\documentclass{aastex6}

\shorttitle{GRB 090510: a S-GRB from a binary NS merger coalescing into a Kerr-Newman BH}
\shortauthors{Ruffini et al.}

\begin{document}

\title{GRB 090510: a genuine short-GRB from a binary neutron star\\coalescing into a Kerr-Newman black hole}

\author{R. Ruffini\altaffilmark{1,2,3,4}, M. Muccino\altaffilmark{1,2}, Y. Aimuratov\altaffilmark{1,3}, C.~L. Bianco\altaffilmark{1,2}, C. Cherubini\altaffilmark{5}, M. Enderli\altaffilmark{1,3}, M. Kovacevic\altaffilmark{1,3}, R. Moradi\altaffilmark{1,2}, A.~V. Penacchioni\altaffilmark{6,7}, G.~B. Pisani\altaffilmark{1,2}, J.~A. Rueda\altaffilmark{1,2,4}, and Y. Wang\altaffilmark{1,2}}

\altaffiltext{1}{Dipartimento di Fisica, Sapienza Universit\`a  di Roma and ICRA, Piazzale Aldo Moro 5, I-00185 Roma, Italy}
\altaffiltext{2}{ICRANet, Piazza della Repubblica 10, I-65122 Pescara, Italy}
\altaffiltext{3}{Universit\'e de Nice Sophia-Antipolis, Grand Ch\^ateau Parc Valrose, Nice, CEDEX 2, France}
\altaffiltext{4}{ICRANet-Rio, Centro Brasileiro de Pesquisas Fisicas, Rua Dr. Xavier Sigaud 150, Rio de Janeiro, RJ, 22290-180, Brazil}
\altaffiltext{5}{ICRA and Unit of Nonlinear Physics and Mathematical Modeling, Department of Engineering, Universit\`a Campus Bio-Medico di Roma, Via Alvaro del Portillo 21, I-00128, Rome, Italy.}
\altaffiltext{6}{University of Siena, Dept. of Physical Sciences, Earth and Environment, Via Roma 56, I-53100 Siena, Italy}
\altaffiltext{7}{ASI Science Data Center, via del Politecnico s.n.c., I-00133 Rome Italy}

\begin{abstract}
In a new classification of merging binary neutron stars (NSs) we separate short gamma-ray bursts (GRBs) in two sub-classes. The ones with $E_{\rm iso}\lesssim10^{52}$~erg coalesce to form a massive NS and are indicated as short gamma-ray flashes (S-GRFs). The hardest, with $E_{\rm iso}\gtrsim10^{52}$~erg, coalesce to form a black hole (BH) and are indicated as genuine short-GRBs (S-GRBs). Within the fireshell model, S-GRBs exhibit three different components: the P-GRB emission, observed at the transparency of a self-accelerating baryon-$e^+e^-$ plasma; the prompt emission, originating from the interaction of the accelerated baryons with the circumburst medium; the high-energy (GeV) emission, observed after the P-GRB and indicating the formation of a BH. GRB 090510 gives the first evidence for the formation of a Kerr BH or, possibly, a Kerr-Newman BH. Its P-GRB spectrum can be fitted by a convolution of thermal spectra whose origin can be traced back to an axially symmetric dyadotorus. A large value of the angular momentum of the newborn BH is consistent with the large energetics of this S-GRB, which reach in the $1$--$10000$~keV range $E_{\rm iso}=(3.95\pm0.21)\times10^{52}$~erg and in the $0.1$--$100$~GeV range $E_{\rm LAT}=(5.78\pm0.60)\times10^{52}$~erg, the most energetic GeV emission ever observed in S-GRBs. The theoretical redshift $z_{\rm th}=0.75\pm0.17$ that we derive from the fireshell theory is consistent with the spectroscopic measurement $z=0.903\pm0.003$, showing the self-consistency of the theoretical approach. All S-GRBs exhibit GeV emission, when inside the \textit{Fermi}-LAT field of view, unlike S-GRFs, which never evidence it. The GeV emission appears to be the discriminant for the formation of a BH in GRBs, confirmed by their observed overall energetics.
\end{abstract}


\section{Introduction}

Thanks to a fortunate coincidence of observations by AGILE, \textit{Fermi}, and \textit{Swift} satellites, together with the optical observations by the VLT/FORS2 and the Nordic Optical Telescope, it has been possible to obtain an unprecedented set of data, extending from the optical-UV, through the X-rays, all the way up to the high energy (GeV) emission, which allowed detailed temporal/spectral analyses on GRB 090510 \citep{Depasquale2010}.

In contrast with this outstanding campaign of observations, a theoretical analysis of the broadband emission of GRB 090510 has been advanced within the synchrotron/self-synchrotron Compton (SSC) and traditional afterglow models \citep[see, e.g., sections 5.2.1 and 5.2.2 in][]{Ackermann2010}.
Paradoxically, this same methodology has been applied in the description of markedly different type of sources: e.g., \citet{Soderberg} for the low energetic long GRB 060218, \citet{2014ApJ...781...37P} for the high energetic long GRB 130427A, and \citet{2006ApJ...650..261S} for the S-GRF 051221A \citep[see also][and references therein]{2008A&A...487..533C}.

In the meantime, it has become evident that GRBs can be subdivided into a variety of classes and sub-classes \citep{Wang2015,Muccino2015,Ruffini2016}, each of them characterized by specific different progenitors which deserve specific theoretical treatments and understanding. 
In addition every sub-class shows different episodes corresponding to specifically different astrophysical processes, which can be identified thanks to specific theoretical treatments and data analysis.
In this article, we take GRB 090510 as a prototype for S-GRBs and perform a new time-resoved spectral analysis, in excellent agreement with the above temporal and spectral analysis performed by, e.g., the \textit{Fermi} team. 
Now this analysis, guided by a theoretical approach successfully tested in this new family of S-GRBs \citep{Muccino2013,Muccino2015}, is directed to identify a precise sequence of different events made possible by the exceptional quality of the data of GRB 090510. 
This include a new structure in the thermal emission of the P-GRB emission, followed by the onset of the GeV emission linked to the BH formation, allowing, as well, to derive the structure of the circumburst medium from the spiky structure of the prompt emission. 
This sequence, for the first time, illustrates the formation process of a BH.

Already in February 1974, soon after the public announcement of the GRB discovery \citep{Strong1975}, \citet{DamourRuffini1975} presented the possible relation of GRBs with the vacuum polarization process around a Kerr-Newman BH.
There, evidence was given for: a) the formation of a vast amount $e^+e^-$-baryon plasma; b) the energetics of GRBs to be of the order of $E_{\rm max}\approx10^{54} M_{\rm BH}/M_\odot$~erg, where $M_{\rm BH}$ is the BH mass; c) additional ultra-high energy cosmic rays with energy up to $\sim10^{20}$~eV originating from such extreme process. A few years later, the role of an $e^+e^-$ plasma of comparable energetics for the origin of GRBs was considered by \citet{CavalloRees} and it took almost thirty years to clarify some of the analogies and differences between these two processes leading, respectively, to the alternative concepts of ``fireball" and ``fireshell" \citep{Aksenov2007,Aksenov2009}. In this article we give the first evidence for the formation of a Kerr Newman BH, in GRB 090510, from the merger of two massive NSs in a binary system.

GRBs are usually separated in two categories, based on their duration properties \citep[e.g.][]{1981Ap&SS..80....3M,Dezalay1992,Klebesadel1992,Kouveliotou1993,Tavani1998}. Short GRBs have a duration $T_{90} \lesssim 2$ s while the remaining ones with $T_{90} \gtrsim 2$ s are traditionally classified as long GRBs. 

Short GRBs are often associated to NS-NS mergers (see e.g.~\citealt{Goodman1986,Paczynski1986,Eichler1989, Narayan1991, Meszaros1997,Rosswog2003,Lee2004,2007PhR...442..166N,2016arXiv160403445E,2016ApJ...824L...6R}; see also \citealt{Berger2014} for a recent review): their host galaxies are of both early- and late-type, their localization with respect to the host galaxy often indicates a large offset \citep{Sahu1997,vanParadijs1997,Bloom2006,Troja2008,Fong2010,Berger2011,Kopac2012} or a location of minimal star-forming activity with typical circumburst medium (CBM) densities of $\sim10^{-5}$--$10^{-4}$ cm$^{-3}$, and no supernovae (SNe) have ever been associated to them. 

The progenitors of long GRBs, on the other hand, have been related to massive stars \citep{WoosleyBloom2006}. However, in spite of the fact that most massive stars are found in binary systems \citep{Smith2014}, that most type Ib/c SNe occur in binary systems \citep{Smith2011} and that SNe associated to long GRBs are indeed of type Ib/c \citep{DellaValle2011}, the effects of binarity on long GRBs have been for a long time largely ignored in the literature. Indeed, until recently, long GRBs have been interpreted as single events in the jetted \textit{collapsar} fireball model (see e.g.~\citealt{Woosley1993,ReesMeszaros1992,Kobayashi1997,Piran2005,Gehrels2009,KumarZhang2015} and references therein). 

Multiple components evidencing the presence of a precise sequence of different astrophysical processes have been found in several long GRBs (e.g.~\citealt{Izzo2012}, \citealt{Penacchioni2012}). Following this discovery, further results led to the introduction of a new paradigm expliciting the role of binary sources as progenitors of the long GRB-SN connection. New developments have led to the formulation of the Induced Gravitational Collapse (IGC) paradigm \citep{Ruffini2001a,Ruffini2007,Rueda2012,Wang2015}. The IGC paradigm explains the GRB-SN connection in terms of the interactions between an evolved carbon-oxygen core (CO$_{\rm core}$) undergoing a SN explosion and its hypercritical accretion on a binary NS companion \citep{Ruffini2015}. The large majority of long bursts is related to SNe and are spatially correlated with bright star-forming regions in their host galaxies \citep{Fruchter2006,Svensson2010} with a typical CBM density of $\sim1$ cm$^{-3}$ \citep{Izzo2012,Penacchioni2012}.

A new situation has occurred with the observation of the high energy GeV emission by the \textit{Fermi}-LAT instrument and its correlation with both long and short bursts with isotropic energy $E_{\mathrm{iso}} \gtrsim 10^{52}$ erg, which has been evidenced in \cite{Wang2015} and \cite{Muccino2015}, respectively. On the basis of this correlation the occurrence of such prolonged GeV emission has been identified with the onset of the formation of a BH \citep{Wang2015,Muccino2015}.

As recalled above, the long GRBs associated to SNe have been linked to the hypercritical accretion process occurring in a tight binary system when the ejecta of an exploding CO$_{\rm core}$ accretes onto a NS binary companion \citep[see, e.g.,][]{Rueda2012,Fryer2014,Becerra}. When the hypercritical accretion occurs in a widely separated system with an orbital separation $>10^{11}$~cm \citep{Becerra}, the accretion is not sufficient to form a BH. For these softer systems with rest-frame spectral peak energy $E_{\rm peak}<200$ keV the upper limit of their observed energy is $E_{\mathrm{iso}}\approx10^{52}$ erg, which corresponds to the maximum energy attainable in the accretion onto a NS \citep{Wang2015}. Such long a burst corresponds to an X-ray flash (XRF). The associated X-ray afterglow is also explainable in terms of the interaction of the prompt emission with the SN ejecta (Fryer et al., in preparation). In these systems no GeV emission is expected in our theory and, indeed, is not observed. Interestingly, a pioneering evidence for such an X-ray flash had already been given in a different context by \cite{Heise2003}, \cite{Amati2004}, and \cite{Soderberg}.
For tighter binaries ($<10^{11}$~cm, \citealt{Becerra}), the hypercritical accretion onto the companion NS leads to the formation of a BH. For these harder systems with $E_{\rm peak}>200$ keV the lower limit of their observed energy is $E_{\mathrm{iso}}\approx10^{52}$ erg, which necessarily needs the accretion process into a BH. An associated prolonged GeV emission occurs after the P-GRB emission and at the beginning of the prompt emission, and originates at the onset of the BH formation \citep{Wang2015}. These more energetic events are referred to as binary-driven hypernovae (BdHNe). Specific constant power-law behaviors are observed in their high energy GeV, X-rays, and optical luminosity light curves \citep{Pisani2013,Ruffini2014,Wang2015}.

In total analogy, the formation of a BH can occur in short bursts, depending on the mass of the merged core of the binary system. When the two NS masses are large enough, the merged core can exceed the NS critical mass and the BH formation is possible. In the opposite case, a massive NS (MNS) is created, possibly, with some additional orbiting material to guarantee the angular momentum conservation. We then naturally expect the existence of two short bursts sub-classes: authentic short GRBs (S-GRBs), characterized by the formation of a BH \citep{Muccino2015}, with $E_{\mathrm{iso}} \gtrsim 10^{52}$ erg, a harder spectrum (see section \ref{epeakeiso}) and associated with a prolonged GeV emission (see section \ref{GeVemission}); short gamma-ray flashes (S-GRFs), producing a MNS \citep{Muccino2015}, with $E_{\mathrm{iso}} \lesssim 10^{52}$ erg. In this second sub-class, of course, the GeV emission should not occur and, indeed, is never observed.

\begin{figure}
\centering
\includegraphics[width=\hsize,clip]{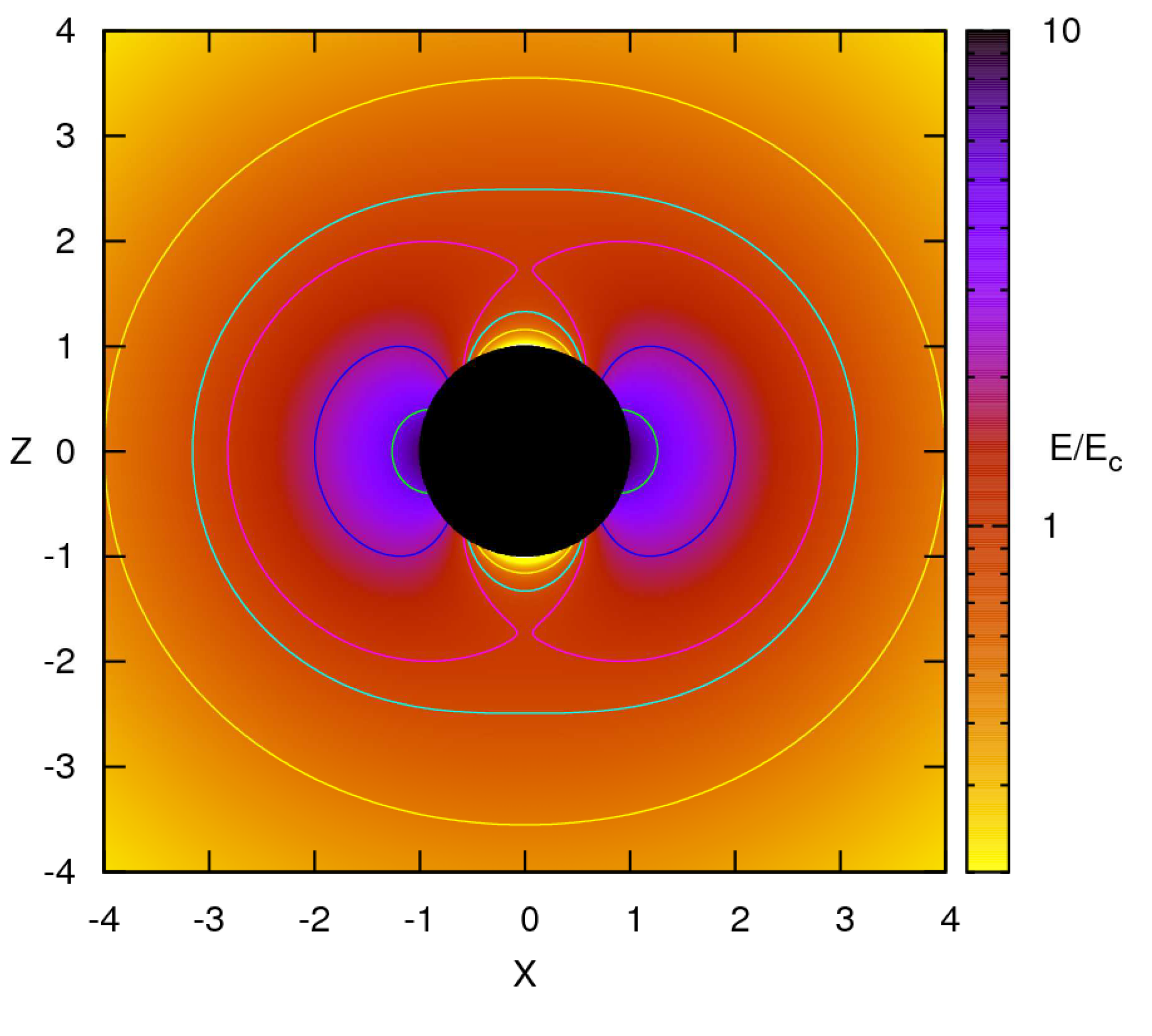}
\caption{Projection of the dyadotorus of a Kerr-Newman BH corresponding to selected values of the ratio $E/E_c$, where $E_c$ is the critical value for vacuum polarization and $E$ is the electric field strength. The plot assumes a black hole mass energy $\mu = M_{\mathrm{BH}} / M_\odot = 10$. Figure reproduced from \cite{Cherubini2009} with their kind permission.}
\label{dyadotorus}
\end{figure}

Following the discovery of the first prototype of this S-GRB class, namely GRB 090227B \citep{Muccino2013}, the first detailed analysis of such a genuine short GRB originating from a binary NS merger leading to a BH was done for GRB 140619B by \cite{Muccino2015}, determining as well the estimated emission of gravitational waves. The latter has been estimated following the method applied by \cite{Oliveira2014} for GRB 090227B. From the spectral analysis of the early $\sim 0.2$ s, they inferred an observed temperature $kT = (324 \pm 33)$ keV of the $e^+e^-$ plasma at transparency (P-GRB), a theoretically derived redshift $z = 2.67 \pm 0.37$, a total burst energy $E^{\mathrm{tot}}_{e^+e^-} = (6.03 \pm 0.79) \times 10^{52}$ erg, a rest-frame peak energy $E_{p,i} = 4.7$ MeV, a baryon load $B = (5.52 \pm 0.73) \times 10^{-5}$, and an average CBM density $n_{\mathrm{CBM}} = (4.7 \pm 1.2) \times 10^{-5}$ cm$^{-3}$. 

We turn in this article to the most interesting case of GRB 090510 which has, in addition to very similar properties of the members of this new class of S-GRB sources, a spectroscopically determined value of the redshift and represents one of the most energetic sources of this family both in the $\gamma$-ray and in the GeV ranges. Actually, a first attempt to analyze GRB 090510 was made by interpreting this source as a long GRB \citep{Muccino2013090510}. An unusually large value of the CBM density was needed in order to fit the data: this interpretation was soon abandoned when it was noticed that GRB 090510 did not fulfill the nesting conditions of the late X-ray emission typical of long GRBs \citep{Ruffini2014}, see also section \ref{xray} and Figure \ref{episode3}.

In light of the recent progress in the understanding of the fireshell theory, we address the interpretation of GRB 090510 as the merging of a binary NS. We give clear evidence for the validity of this interpretation. In view of the good quality of the data both in $\gamma$- rays and in the GeV range, we have performed a more accurate description of the P-GRB, best fitted by a convolution of thermal spectra. This novel feature gives the first indication for the existence of an axially symmetric configuration of the dyadotorus emitting the $e^+e^-$ plasma which had been previously theoretically considered and attentively searched for. This gives the first indication that indeed the angular momentum plays a role and a dyadotorus is formed, as theoretically predicted in a series of papers \citep[see][and Figure \ref{dyadotorus}]{Cherubini2009,Ruffini2009}. 
This naturally leads to the evidence for the formation of a rotating BH as the outcome of the gravitational collapse. We turn then to the main new feature of GRB 090510 which is the high energy $0.1$--$100$ GeV emission (see Figure \ref{090510GeV}). The direct comparison of the GeV emission in this source and in the BdHNe 130427A shows the remarkable similarities of these two GeV components (see Figure \ref{090510GeV}).
The fact that the S-GRB 090510 originates from a binary NS merger and the BdHN 130427A from the IGC of a SN hypercritical accretion process onto a companion NS clearly points to the BH as originating this GeV emission, the reason being that these two astrophysical systems are different in their progenitors and physical process and have in the formation of a BH their unique commonality.

This paper is structured as follows: in section 2 we summarize the relevant aspects of the fireshell theory and compare and contrast it with alternative approaches. In section 3 we discuss the recent progress on the NS equilibrium configuration relevant for S-GRBs and BdHNe.
In section 4 we move on to describe the observations of GRB 090510 and their analysis. The S-GRB nature of GRB 090510 is justified in section 5, and we offer an interpretation of our results in section 6. 
Section 7 concludes this work.

A standard flat ${\Lambda}$CDM cosmological model with ${\Omega}_m = 0.27$ and $H_0 = 71$ km s$^{-1}$ Mpc$^{-1}$ is adopted throughout the paper.

\section{Summary of the fireshell model}

The fireshell scenario \citep{Ruffini2001a,Ruffini2001b,Ruffini2001c}, has been initially introduced to describe a GRB originating in a gravitational collapse leading to the formation of a Kerr-Newman BH. A distinct sequence of physical and astrophysical events are taken into account:
\begin{itemize}
\item[1)] An optically thick pair plasma -- the fireshell of total energy $E_{e^+e^-}^{\mathrm{tot}}$ -- is considered. As a result, it starts to expand and accelerate under its own internal pressure \citep{Ruffini1999}. The baryonic remnant of the collapsed object is engulfed by the fireshell -- the baryonic contamination is quantified by the baryon load $B = M_Bc^2 / E_{e^+e^-}^{\mathrm{tot}}$ where $M_B$ is the mass of the baryonic remnant \citep{Ruffini2000,Aksenov2007,Aksenov2009}.
\item[2)] After the engulfment, the fireshell is still optically thick and continues to self-accelerate until it becomes transparent. When the fireshell reaches transparency, a flash of thermal radiation termed Proper-GRB (P-GRB) is emitted \citep{Ruffini1999,Ruffini2000}.
\item[3)] In GRBs, the $e^+e^-$-baryon plasma evolves from the ultra-relativistic region near the BH all the way reaching ultra-relativistic velocities at large distances. To describe such a dynamics which deals with unprecedentedly large Lorentz factors and also regimes sharply varying with time, in \citet{Ruffini2001a} it has been introduced the appropriate relative spacetime transformation paradigm. This paradigm gives particular attention to the constitutive equations relating four time variables: the comoving time, the laboratory time, the arrival time, and the arrival time at the detector corrected by the cosmological effects. This paradigm is essential for the interpretation of the GRB data: the absence of adopting such a relativistic paradigm in some current works has led to a serious misinterpretation of the GRB phenomenon.
\item[4)] In compliance with the previous paradigm, the interactions between the ultra-relativistic shell of accelerated baryons left over after transparency and the CBM have been considered. They lead to a modified blackbody spectrum in the co-moving frame \citep{Patricelli2012}. The observed spectrum is however non-thermal in general; this is due to the fact that, once the constant arrival time effect is taken into account in the EQuiTemporal Surfaces \citep[EQTS, see][]{Bianco2005a,Bianco2005b}, the observed spectral shape results from the convolution of a large number of modified thermal spectra with different Lorentz factors and temperatures.
\item[5)] All the above relativistic effects, after the P-GRB emission, are necessary for the description of the prompt emission of GRBs, as outlined in \citet{Ruffini2001b}. The prompt emission originates in the collisions of the accelerated baryons, moving at Lorentz factor $\gamma\approx100$--$1000$, with interstellar clouds of CBM with masses of $\sim10^{22}$--$10^{24}$~g, densities of $\sim0.1$--$1$~cm$^{-3}$ and size of $\sim10^{15}$--$10^{16}$~cm, at typical distances from the BH of $\sim10^{16}$--$10^{17}$~cm (see, e.g., \citealt{Izzo2012} for long bursts). Our approach differs from alternative tratments purporting late activities from the central engine (see, e.g., the \textit{collapsar} model in \citealt{Woosley1993}, \citealt{1999ApJ...518..356P}, \citealt{WoosleyBloom2006} and references therein, and the \textit{magnetar} model in \citealt{2001ApJ...552L..35Z}, \citealt{2006Sci...311.1127D}, \citealt{2011MNRAS.413.2031M}, \citealt{2012MNRAS.419.1537B}, \citealt{2014ApJ...785...74L}, and references therein).
\item[6)] $E_{e^+e^-}^{\mathrm{tot}}$ and $B$ are the only two parameters that are needed in a spherically symmetric fireshell model to determine the physics of the fireshell evolution until the transparency condition is fulfilled. Three additional parameters, all related to the properties of the CBM, are needed to reproduce a GRB light curve and its spectrum: the CBM density profile $n_{\mathrm{CBM}}$, the filling factor $\mathcal{R}$ that accounts for the size of the effective emitting area, and an index $\alpha$ that accounts for the modification of the low-energy part of the thermal spectrum \citep{Patricelli2012}. They are obtained by running a trial-and-error simulation of the observed light curves and spectra that starts at the fireshell transparency.
\item[7)] A more detailed analysis of pair cration process around a Kerr-Newman BH has led to the concept of dyadotorus \citep{Cherubini2009}. There, the axially symmetric configuration with a specific distribution of the $e^+e^-$, as well as its electromagnetic field, have been presented as function of the polar angle. The total spectrum at the transparency of the $e^+e^-$plasma is a convolution of thermal spectra at different angles.
\end{itemize}

This formalism describing the evolution of a baryon-loaded pair plasma is describable in terms of only three intrinsic parameters: the $e^+e^-$ plasma energy $E_{e^+e^-}^{\rm tot}$, the baryon load $B$, and the specific angular momentum $a$ of the incipient newly-formed BH. It is, therefore, independent of the way the pair plasma is created. 

\begin{figure}
\centering
\includegraphics[width=\hsize]{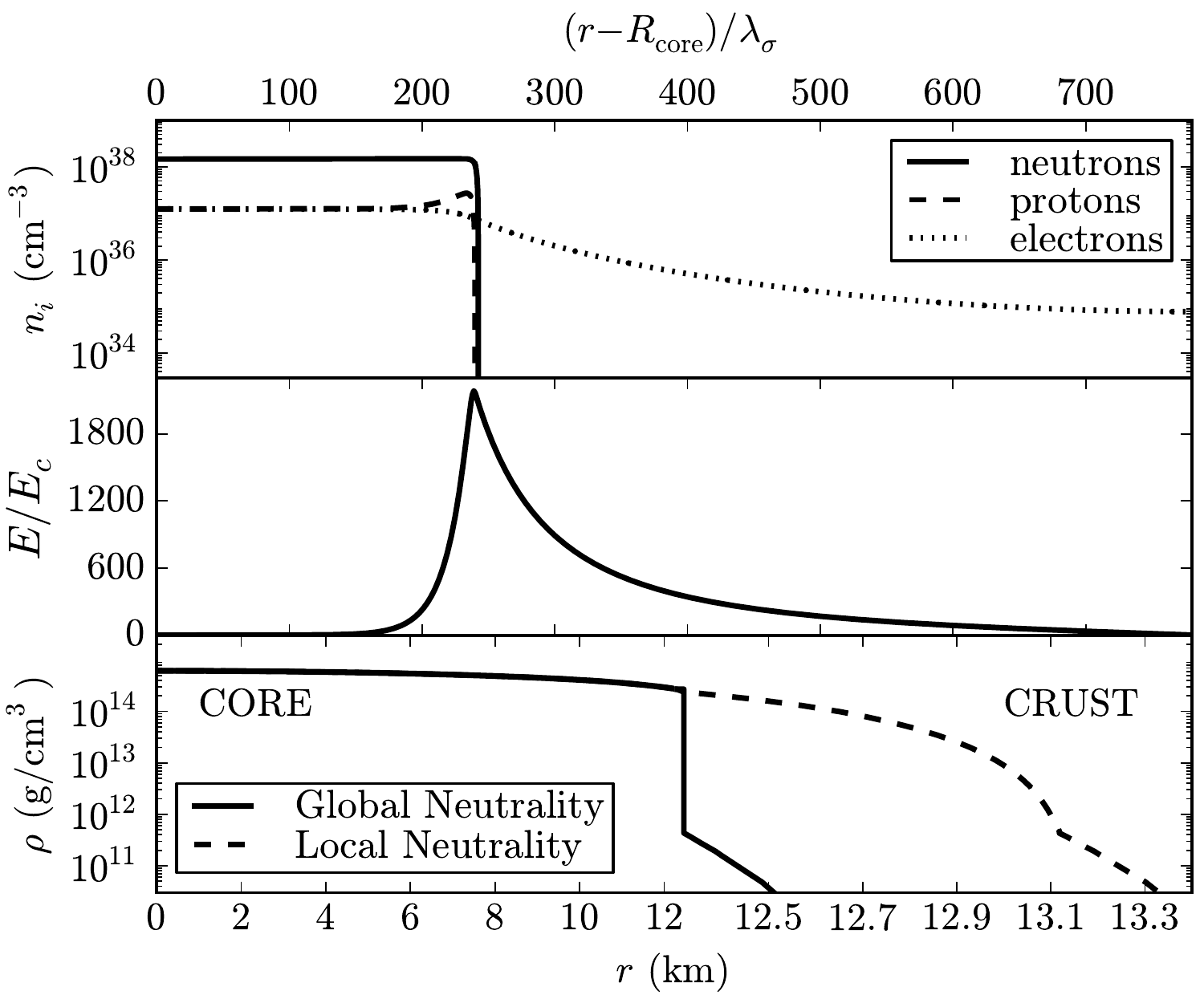}
\caption{The particle density profiles (\textit{upper panel}) and the electric field in units of $E_c$ (\textit{middle panel}) in the core-crust transition layer normalized to the $\sigma$-meson Compton wavelength $\lambda_\sigma=\hbar/(m_\sigma c)\sim0.4$~fm. \textit{Lower panel}: density profile inside a NS star with central density $\rho\sim5\rho_{\rm nuc}$, where $\rho_{\rm nuc}$ is the nuclear density, from the solution of the TOV equations (locally neutral case) and the globally neutral solution presented in \cite{Belvedere2012}. The density at the edge of the crust is the neutron drip density $\rho_{\rm drip}=4.3\times10^{11}$~g~cm$^{−3}$. Reproduced from \cite{Belvedere2012} with their kind permission.}
\label{Nanda}
\end{figure}

In addition to the specific case, developed for the sake of example, of the dyadotorus created by a vacuum polarization process in an already formed Kerr-Newman BH, more possibilities have been envisaged in the meantime:
\begin{itemize}
\item[a)] The concept of dyadotorus can be applied as well in the case of a pair plasma created via the $\nu \bar{\nu} \leftrightarrow e^+e^-$ mechanism in a NS merger as described in \cite{Narayan1992}, \cite{SalmonsonWilson2002}, \cite{Rosswog2003}, \cite{2011MNRAS.410.2302Z}, assuming that the created pair plasma is optically thick. The relative role of neutrino and weak interactions vs. the electromagnetic interactions in building the dyadotorus is currently topic of intense research.
\item[b)] Equally important are the relativistic magneto-hydrodynamical process leading to a dyadotorus, indicated in the general treatment of \cite{RRWilson1975}, and leading to the birth of a Kerr-Newman BH, surrounded by an opposite charged magnetosphere in a system endowed with global charge neutrality. Active research is ongoing.
\item[c)] Progress in understanding the NS equilibrium configuration imposing the global charge neutrality condition, as opposed to the local charge neutrality usually assumed \citep{2011PhRvC..83d5805R,2011PhLB..701..667R,2011NuPhA.872..286R,2013IJMPD..2260007R,2014PhRvC..89c5804R}. A critical mass for a non-rotating NS $M_{\mathrm{crit}}^{\mathrm{NS}} \approx 2.67 M_\odot$ has been found for the NL3 nuclear equation of state \citep{Belvedere2012}. The effects of rotation and of the nuclear equation of state on the critical mass is presented in \cite{Belvedere2014,Belvedere2015} and in \cite{Cipolletta2015}.
The existence of electromagnetic fields close to the critical value has been evidenced in the interface between the core and the crust in the above global neutrality model, as well as very different density distributions in the crust and in the core, which could play an important role during the NS--NS mergers \citep[see Figure~\ref{Nanda} and][]{Oliveira2014}.
\end{itemize}

The above three possibilities have been developed in recent years, but they do not have to be considered exaustive for the formation of a dyadotorus endowed by the above three parameters.

In conclusion the evolution in the understanding of the GRB phenomenon, occurring under very different initial conditions, has evidenced the possibility of using the
dyadotorus concept for describing sources of an optically thick baryon-loaded $e^+e^-$ plasma within the fireshell treatment in total generality.

\section{On the role of $\approx10^{52}$~erg limit for S-GRBs and BdHNe}

The key role neutrino emission in the hypercritical accretion process onto a NS has been already examined in the literature \citep[see, e.g.,][]{Zeldovich1972,RRWilson1973}. 
The problem of hypercritical accretion in a binary system composed of a CO$_{\rm core}$ and a companion NS has been studied in \citet{Becerra,2016arXiv160602523B} (see also references therein). 
The energy released during the process, in form of neutrinos and photons, is given by the gain of gravitational potential energy of the matter being accreted by the NS and depends also on the change of binding energy of the NS while accreting both matter and on the angular momentum carried by the accreting material (see, e.g., \citealt{2016arXiv160602523B} and \citealt{Ruffini2016}).
For a typical NS mass of $\approx1.4$~M$_\odot$, a value observed in galactic NS binaries \citep{2011A&A...527A..83Z,2014arXiv1407.3404A}, and a NS critical mass $M_{\rm crit}^{\rm NS}$ in the range from $2.2$~M$_\odot$ up to $3.4~M_\odot$ depending on the equations of state and angular momentum \citep[see][for details]{Becerra,2016arXiv160602523B,Cipolletta2015}, the accretion luminosity can be as high as $L_{\rm acc}\sim 0.1 \dot{M_b} c^2\sim 10^{47}$--$10^{51}$~erg~s$^{-1}$ for accretion rates $\dot{M_b}\sim 10^{-6}$--$10^{-2}~M_\odot$~s$^{-1}$ \citep[see][for details]{Becerra,2016arXiv160602523B}. 
For binary systems with a separation $\sim 10^{10}$~cm ($P\sim 5$~min), our numerical simulations indicate that: a) the accretion process duration lasts $\Delta t_{\rm acc}\sim 10^2$~s \citep[see, e.g.,][]{Becerra,2016arXiv160602523B}, b) the NS collapses to a BH, and c) a total energy larger than $\approx10^{52}$~erg is released during the hypercritical accretion process. These systems correspond to the BdHNe \citep{2016arXiv160602523B}.
For systems with larger separations the hypercritical accretion is not sufficient to induce the collapse of the NS into a BH and the value of $\approx10^{52}$~erg represents a theoretical estimate of the upper limit to the energy emitted by norm in the hypercritical accretion process. This sub-class of sources corresponds to the XRFs \citep{2016arXiv160602523B}.

\begin{figure}
\centering
\includegraphics[width=0.9\hsize]{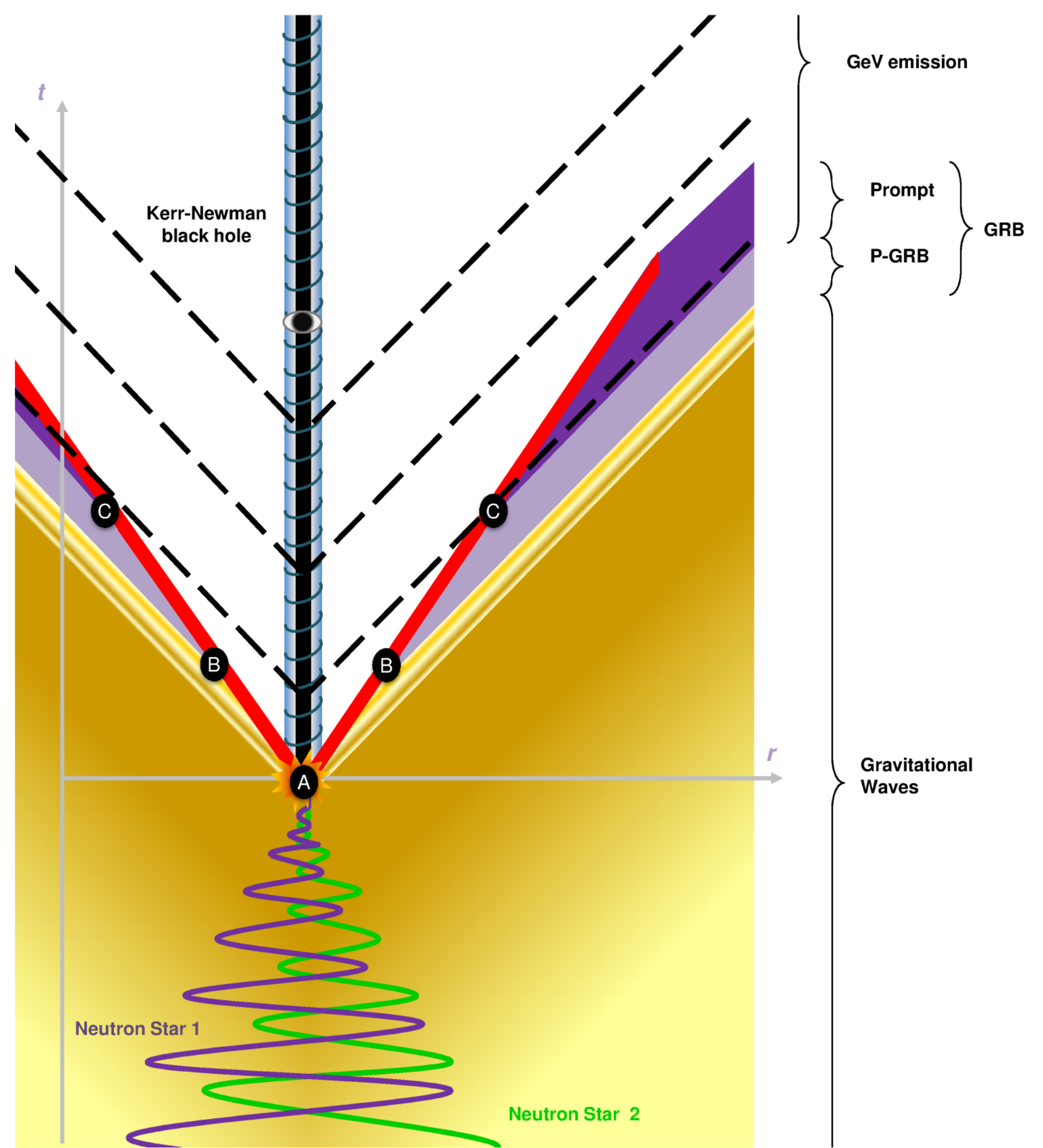}
\caption{Space-time diagram of a S-GRB, a binary NS merger leading to BH formation (taken from Enderli et al. 2015 with their kind premission \footnote{\url{ http://pos.sissa.it/archive/conferences/233/073/SWIFT\%2010_073.pdf}}). The binary orbit gradually shrinks due to energy loss through gravitational waves emission (yellow-brown). At point A, the merger occurs: the fireshell (in red) is created and starts its expansion. It reaches transparency at point B, emitting the P-GRB (light purple). The prompt emission (deep purple) then follows at point C. The dashed lines represent the GeV emission (delayed relative to the start of the GRB) originating in the newly-born BH.
This space-time diagram well illustrates how the GeV emission originates in the newly-born BH and follows a different space-time path from the prompt emission, contrary to what stated in \citet{Ackermann2010}. The prompt emission originates from the interactions of the baryons, accelerated to ultrarelativistic Lorentz factors during the pair-baryon electromagnetic pulse, with the clumpy circumburst medium (see section 2). The analysis of the spiky structure of the prompt emission allows to infer the structure of the circumburst medium (see Figure~\ref{090510simlc}). There is the distinct possibility that the GeV emission prior to $0.6$ s in the arrival time may interact with the prompt emission. In this sense the work by \citet{2011ApJ...726L...2Z} may become of interest.}
\label{rt_gw}
\end{figure}

The same energetic considerations do apply in the analysis of the hypercritical accretion occurring in a close binary NS system undergoing merging \citep{Ruffini2015}.
Therefore, in total generality, we can conclude that the energy emitted during a NS--NS merger leading to the formation of a BH should be larger than $\approx10^{52}$~erg (see Figure \ref{rt_gw}).

The limit of $\approx10^{52}$~erg clearly depends on the initial NS mass undergoing accretion, by norm assumed to be $\approx1.4$~M$_\odot$, and on the yet unknown value of $M_{\rm crit}^{\rm NS}$, for which only an absolute upper limit of $3.2$~M$_\odot$ has been established for the non-rotating case \citep{Rhoades1974}.
As already pointed out in \citet{Ruffini2015}, for NS--NS mergers, the direct determination of the energy threshold  of $\approx10^{52}$~erg dividing S-GRFs and S-GRBs, as well as XRFs and BdHNe, provides fundamental informations for the determination of the actual value of $M_{\mathrm{crit}}^{\mathrm{NS}}$, for the minimum mass of the newly-born BH, and for the mass of the accreting NS.

\section{Analysis of GRB 090510}

In this section, we summarize the observations of GRB 090510 as well as the data analysis. We used \emph{Fermi} (GBM and LAT) and \emph{Swift}/XRT data for the purposes of this work.

\subsection{Observations}

The \emph{Fermi}/GBM instrument \citep{Meegan2009} was triggered at $T_0=$00:22:59.97 UT on May 10, 2009 by the short and bright burst GRB 090510 (\citealt{Guiriec2009}, trigger 263607781 / 090510016). The trigger was set off by a precursor emission of duration 30 ms, followed $\sim $ 0.4 s later by a hard episode lasting $\sim 1$ s. This GRB was also detected by Swift \citep{Hoversten2009}, \emph{Fermi}/LAT \citep{OhnoPelassa2009}, AGILE \citep{Longo2009}, Konus-Wind \citep{Golenetskii2009}, and Suzaku-WAM \citep{Ohmori2009}. The position given by the GBM is consistent with that deduced from \emph{Swift} and LAT observations. 

During the first second after LAT trigger at 00:23:01.22 UT, \emph{Fermi}/LAT detected over 50 events (respectively over 10) with an energy above 100 MeV (respectively above 1 GeV) up to the GeV range, and more than 150 (respectively more than 20) within the first minute \citep{Omodei2009}. This makes GRB 090510 the first bright short GRB with an emission detected from the keV to the GeV range.

Observations of the host galaxy of GRB 090510, located by VLT/FORS2, provided a measurement of spectral emission lines. This led to the determination of a redshift $z = 0.903 \pm 0.003$ \citep{Rau2009}. The refined position of GRB 090510 obtained from the Nordic Optical Telescope \citep{Olofsson2009} is offset by 0.7" relative to the center of the host galaxy in the VLT/FORS2 image. At $z = 0.903$, this corresponds to a projected distance of 5.5 kpc. The identified host galaxy is a late-type galaxy of stellar mass $5 \times 10^{9} \ M_\odot$, with a rather low star-forming rate SFR $ = 0.3\ M_\odot \ \mathrm{yr}^{-1}$ (\citealt{Berger2014} and references therein).

\subsection{Data analysis}

\begin{figure}
\centering
\includegraphics[width=0.9\hsize]{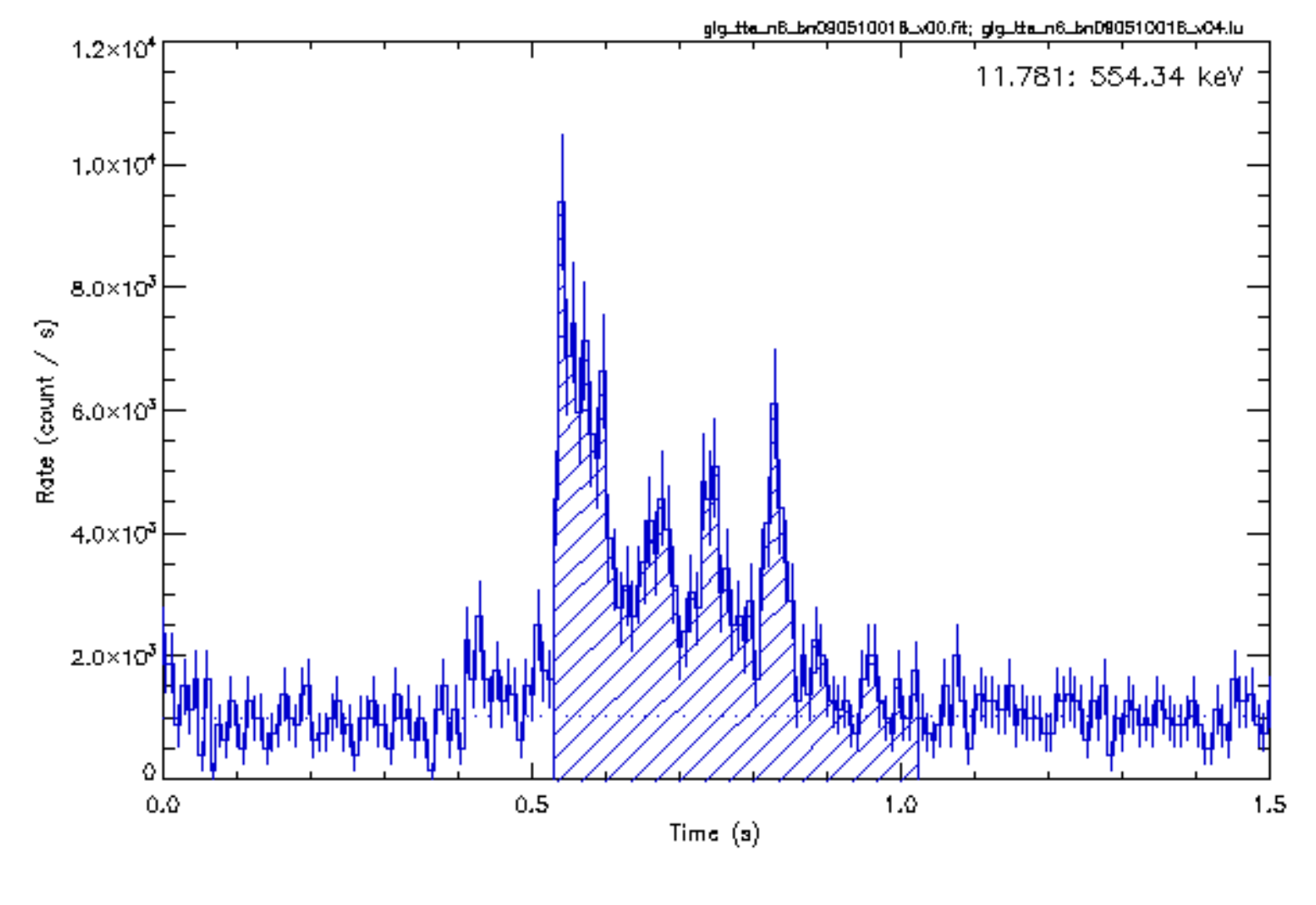}
\includegraphics[width=0.9\hsize]{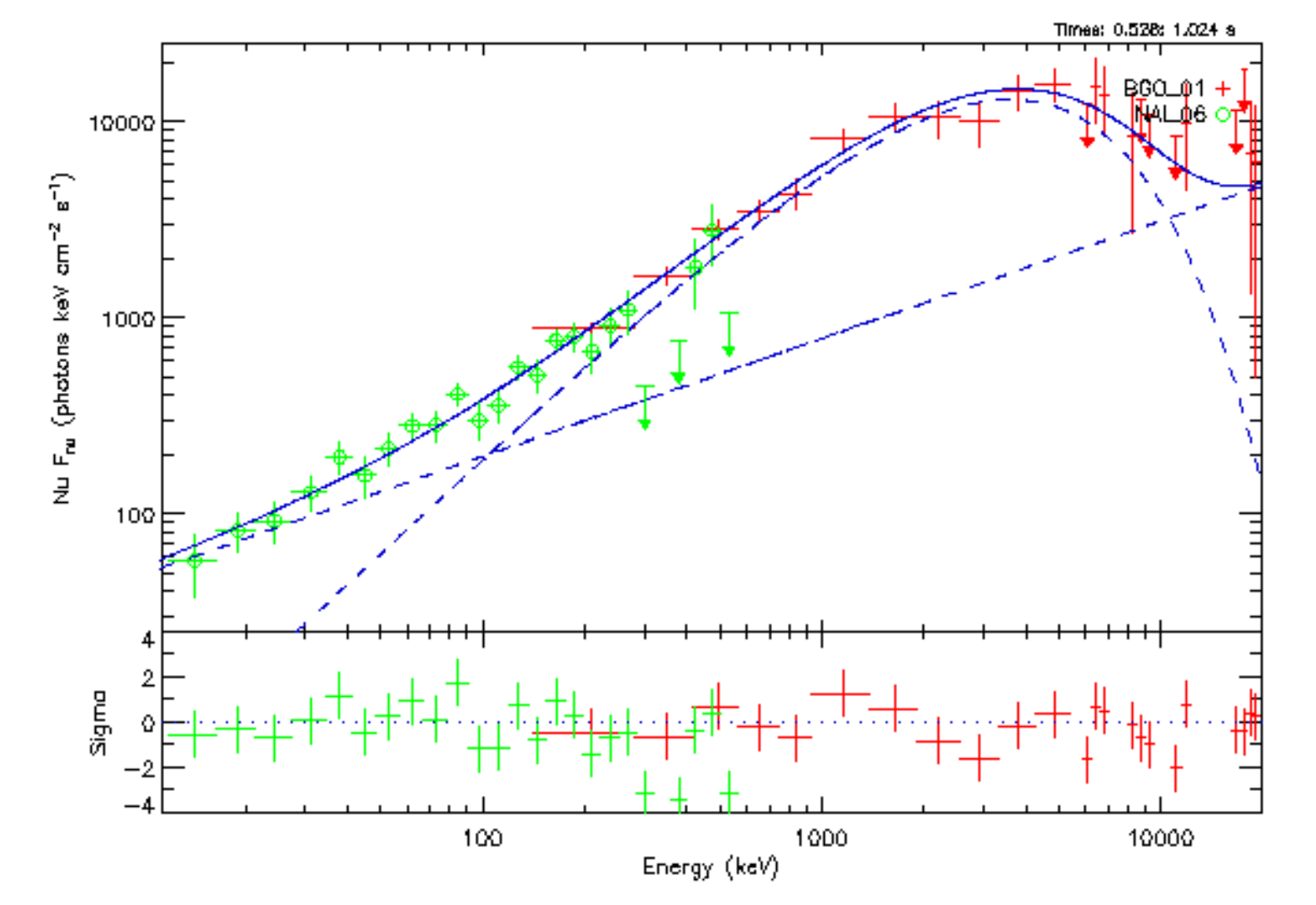}
\caption{Upper panel: GBM NaI-n6 light curve of GRB 090510 and interval considered to compute $E_{\mathrm{iso}}$. Lower panel: Comptonized + power law best fit of the corresponding spectrum (from $T_0+0.528$ to $T_0+1.024$ s).}
\label{spectotal}
\end{figure}

\begin{table*}
\centering
\begin{tabular}{lccccccc}
\hline\hline
  & Model 	& $ \mathrm{C-STAT} / \mathrm{DOF}$ & $E_{\mathrm{peak}}$ (keV)	& $\alpha$	& $\beta$ 	& $\gamma$	& $kT$ (keV)\\
 \hline
  & Band		& 221.46 / 237 & $2987 \pm 343$ & $-0.64 \pm 0.05$ & $-3.13 \pm 0.42$ & -- & --\\
  & Comp	 	& 392.65 / 238 & $3020 \pm 246$ & $-0.64 \pm 0.05$ & -- & -- & --	\\
  & Comp + PL 	& 209.26 / 236 & $2552 \pm 233$ & $-0.26 \pm 0.14$ & -- & $-1.45 \pm 0.07$ & --\\
  & PL 			& 492.83 / 239 & -- & -- & -- & $-1.20 \pm 0.02$ & --\\
  & BB + PL		 & 250.09 / 237& -- & -- & -- & $-1.38 \pm 0.04$ & $477.5 \pm 24.9$\\
 \hline
 \end{tabular}
\caption{Fitting parameters of the time interval $T_0+0.528$ s to $T_0+0.640$ s identified with the P-GRB, where $\alpha$ is the low-energy index of the Comp or Band component, $\beta$ is the high-energy index of the Band component, $E_{\mathrm{peak}}$ is the peak energy of the Comp or Band component, $\gamma$ is the power law index, and $kT$ is the temperature of the blackbody component.}
\label{tab:fit}
\end{table*}

Our analysis focused on \emph{Fermi} (GBM and LAT) and \emph{Swift}/XRT data. The \emph{Fermi}/GBM signal is the most luminous in the NaI-n6 ($8$--$900$ keV, dropping the overflow high-energy channels and cutting out the K-edge between $\sim 30$ and $\sim 40$ keV) and BGO-b1 ($260$ keV -- $40$ MeV, again dropping the overflow high-energy channels) detectors. We additionally considered \emph{Fermi}/LAT data in the $100$ MeV -- $100$ GeV energy range. We made use of standard software in our analysis: GBM time-tagged data -- suitable in particular for short GRBs -- were analyzed with the \url{rmfit} package\footnote{\url{http://fermi.gsfc.nasa.gov/ssc/data/analysis//rmfit/vc\_rmfit\_tutorial.pdf}}; LAT data were analyzed with the Fermi Science tools\footnote{\url{ http://fermi.gsfc.nasa.gov/ssc/data/analysis//documentation/Cicerone/}}. The data were retrieved from the Fermi science support center\footnote{\url{http://fermi.gsfc.nasa.gov/ssc/data/access/}}. \emph{Swift}/XRT data were retrieved from the UK Swift Data Centre at the University of Leicester\footnote{\url{http://www.swift.ac.uk/archive/index.php}} and they have been reduced and analyzed using XSPEC.

Using GBM time-tagged event data binned in $16$ ms intervals, the best fit in the interval $T_0 + 0.528$ s to $T_0 +1.024$ s is a Comptonized + power law model (see Figure \ref{spectotal}). Using this spectral model we find an isotropic energy $E_{iso} = (3.95 \pm 0.21) \times 10^{52}$ erg. The observed peak energy of the best-fit Band model of the time-integrated GBM data is $4.1\pm0.4$ MeV, which corresponds to a rest-frame value of $7.89\pm0.76$ MeV.

The best-fit model during the first pulse (from $T_0 + 0.528$ s to $T_0 + 0.640$ s) in the 8 keV -- 40 MeV range is also a Comptonized + power law, preferred over a power law (PL, $\Delta \mathrm{C-STAT} = 100$), a blackbody plus PL (BB + PL, $\Delta \mathrm{C-STAT} = 41$), or a Band model ($\Delta \mathrm{C-STAT} = 12$). The fitting statistics are summarized in Table \ref{tab:fit}. The peak energy $E_{\mathrm{peak}}$ of the Comptonized component is $2.6$ MeV. The total isotropic energy contained in this time interval is $\sim 1.77 \times 10^{52}$ erg, while the isotropic energy contained in the Comptonized part reaches $\sim 1.66 \times 10^{52}$ erg.

\section{GRB 090510 as a S-GRB}

We here justify the interpretation of GRB 090510 as a S-GRB event. In addition to the duration and hardness properties that are similar to other GRBs interpreted as binary NS mergers, the pattern of the late X-ray emission and the position of GRB 090510 in the $E_{\mathrm{peak}} - E_{\mathrm{iso}}$ plane favor this interpretation.

\subsection{Late X-ray emission (Episode 3)} \label{xray}

An important feature of BdHNe is the existence of a pattern in the behavior of their $0.3$--$10$ keV late X-ray luminosity light curves that we refer to as the Episode 3 (see e.g.~\citealt{Wang2015}). This emission is observationally characterized by the overlapping of the common late power-law behavior \citep{Pisani2013} as well as by the nesting, namenly an inverse (direct) proportionality relation between the duration (the luminosity) of the plateau phase and the energy of the GRB emission: the more energetic the source, the smaller (higher) the duration (the luminosity) of the plateau \citep{Ruffini2014}.

If GRB 090510 were to be an IGC event exploding in a high-density environment, this characteristic Episode 3 would be expected and should be seen. Thanks to adequate coverage by the \emph{Swift}/XRT instrument, the late X-ray ($0.3$--$10$ keV) emission of GRB 090510 has been well sampled. We computed its rest-frame $0.3$--$10$ keV luminosity light curve, using a simple power law spectral fit and taking care of the K-correction as follows:
\begin{equation}
L_{\mathrm{rf}} = 4 \pi d_l^2(z) f_{\mathrm{obs}} \frac{\int^{10 / (1+z)\ \mathrm{ keV}}_{0.3/(1+z)\ \mathrm{ keV}} E^{-\gamma} dE}{\int^{10\ \mathrm{keV}}_{0.3\ \mathrm{keV}} E^{-\gamma} dE}
\end{equation}
where $f_{\mathrm{obs}}$ is the XRT flux (in erg s$^{-1}$ cm$^{-2}$) in the observed $0.3$--$10$ keV range, $d_l$ is the luminosity distance, $\gamma$ is the photon index of the XRT spectrum, and $E^{-\gamma}$ is the spectral model (here, a simple power law) fitting the observed XRT flux.

\begin{figure}
\centering
\includegraphics[width=\hsize]{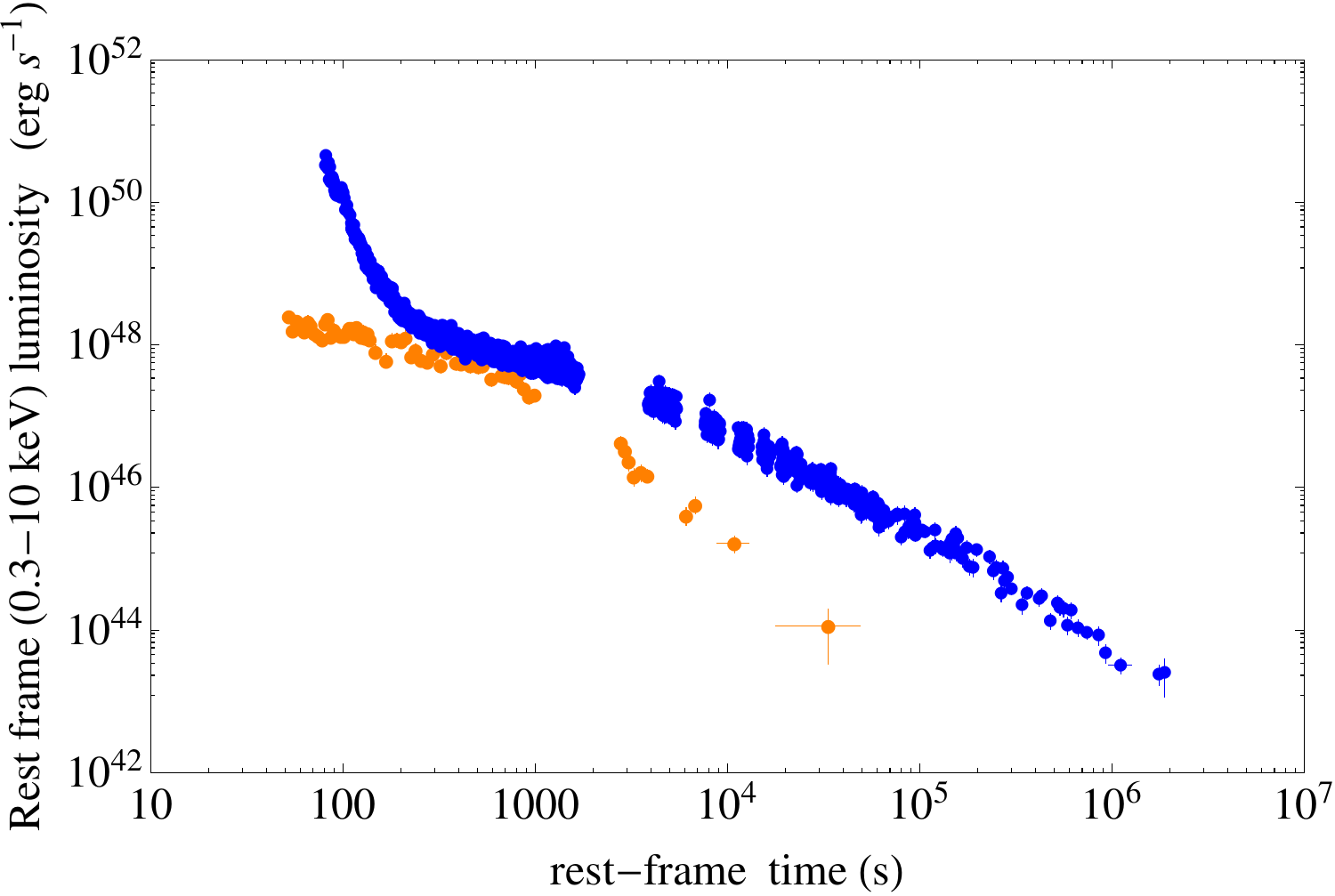}
\caption{Rest-frame $0.3$--$10$ keV luminosity light curves of GRB 090510 (in orange) and GRB 090618 (in blue), the prototypical IGC source. An overlapping pattern has been observed in IGC sources \citep{Pisani2013}, as well as a nesting behavior \citep{Ruffini2014}; it is clear from the deviation between the two light curves that GRB 090510 does not follow this characteristic pattern, thereby confirming its non-IGC nature.}
\label{episode3}
\end{figure}

The rest-frame $0.3$--$10$ keV luminosity light curve is plotted in Figure \ref{episode3}: the comparison with the prototypical IGC source GRB 090618 shows a clear deviation from the overlapping and nesting patterns. Indeed, the late X-ray emission of GRB 090510 is much weaker than that of typical IGC sources, and does not follow the typical power law behavior as a function of time which has a slope $- 1.7 \leq \alpha_X \leq - 1.3$. As a consequence, this result is inconsistent with the hypothesis of GRB 090510 being a BdHN disguised as a short burst. Instead, the interpretation as a S-GRB is in full agreement with the theory and the data (see below).

\subsection{$E_{\mathrm{peak}} - E_{\mathrm{iso}}$ relation} \label{epeakeiso}

\begin{figure}
\centering
\vspace{-0.4cm}
\includegraphics[angle=270,width=\hsize]{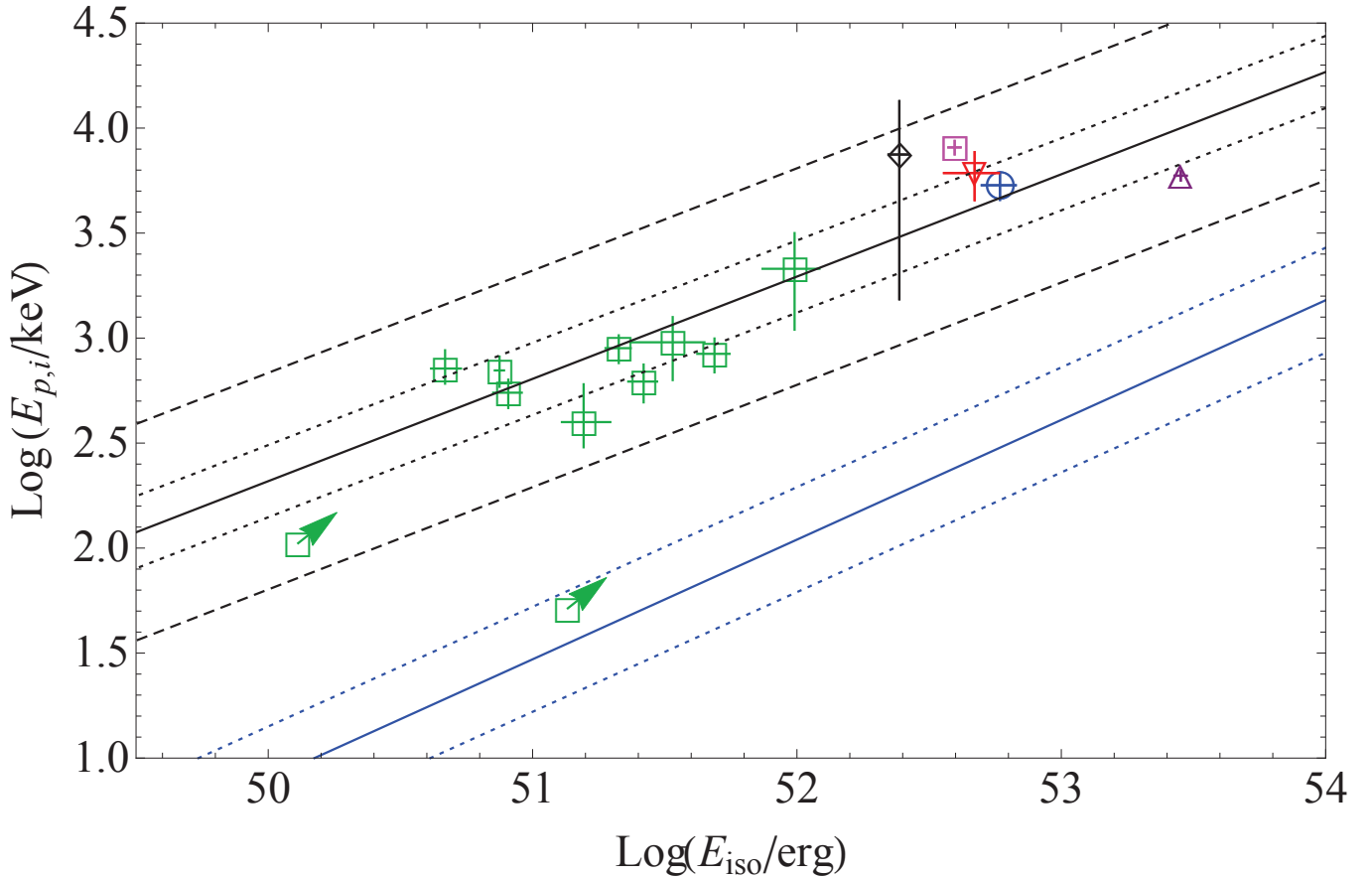}
 \vspace{-0.4cm}
\caption{$E_{\mathrm{peak}} - E_{\mathrm{iso}}$ plot of all short bursts with redshift. The black line marks the relation for short GRBs (which includes the theoretical redshifts we obtained for 4 GRBs). This relation takes the form $\log E_{\mathrm{peak}} = A + \gamma (\log E_{\mathrm{iso}})$ where $A = -22.0 \pm 3.2$, $\gamma = 0.49 \pm 0.06$, and $E_{\mathrm{peak}}$ and $E_{\mathrm{iso}}$ are respectively given in keV and erg. The dotted and dashed lines represent respectively the $1\sigma$ and $3\sigma$ scatter of the relation ($\sigma_{\mathrm{sc}} = 0.17 \pm 0.04$ dex). 
Green boxes indicate S-GRFs with a measured redshift; only lower limits are available for the two S-GRFs singled out by an arrow. GRB 090510 is marked by the pink square. The other four symbols indicate S-GRBs with a redshift derived from the fireshell analysis. The black diamond indicates GRB 081024B, the red inverted triangle GRB 140402A, the blue square GRB 140619B, and the purple triangle GRB 090227B. 
For comparison, the blue line marks the relation for long GRBs given in \cite{Calderone2015} $\log E_{\mathrm{peak}} = A + \gamma (\log E_{\mathrm{iso}} - B)$ where $A = 2.73$, $B = 53.21$, $\gamma =0.57 \pm 0.06$. The dotted lines represent the $1\sigma$ scatter of the relation ($\sigma_{\mathrm{sc}} = 0.25$ dex).}
\label{calderone}
\end{figure}

Although the sample of short bursts with a measured redshift and an estimate of $E_{\mathrm{peak}}$ is of modest size in comparison to that of long GRBs, it has been noted that a relation similar to the Amati one \citep{Amati2002,2013IJMPD..2230028A} exists for short bursts (see e.g.~\citealt{Zhang2012,Calderone2015}). Plotted in Figure \ref{calderone}, this $E_{\mathrm{peak}} - E_{\mathrm{iso}}$ relation has almost the same slope of the Amati relation but drastically differs in their amplitudes.

While \cite{Zhang2012} extended this analysis to the above defined S-GRFs, we have recently added four S-GRBs in this  $E_{\mathrm{peak}} - E_{\mathrm{iso}}$ relation, which we have called the MuRuWaZha relation \citep{RuffiniGCN}.

With the parameters $E_{\mathrm{peak}} = (7.89 \pm 0.76)$ MeV and $E_{\mathrm{iso}} = (3.95 \pm 0.21) \times 10^{52}$ erg obtained in the previous sections, GRB 090510 falls right on the relation fulfilled by S-GRBs, and far from that of long GRBs (see Figure \ref{calderone}). 
This point further strengthens the identification of GRB 090510 as a S-GRB.

\subsection{The offset from the host galaxy}

Long bursts are known to trace star formation (e.g.~\citealt{Bloom2002}). They explode mainly in low-mass galaxies with high specific star-formation rates. On the other hand, short bursts occur in a wider range of host galaxies, including old, elliptical galaxies with little star formation and young galaxies. Their median projected offset from the center of their host, about $5$ kpc, is also known to be four times larger than that of long bursts \citep{Bloom2002}. With a projected offset of $5.5$ kpc, as detailed previously, GRB 090510 falls in the typical short bursts range. Its host galaxy is a late-type one.

The results of the fireshell analysis summarized in the next section also support this conclusion. The average CBM density of GRB 090510 is indeed evaluated at $\langle n_{\mathrm{CBM}}\rangle = 8.7 \times 10^{-6}\ \mathrm{ cm}^{-3}$ (see next section), a low value that is typical of galactic halos.

\section{Interpretation}

GRB 090510 exhibits several peculiar features: the spectrum of the P-GRB is not purely thermal; a weak precursor emission is clearly seen; a GeV emission is observed -- which never occurs in S-GRFs but appears to be a general property of the S-GRBs. This section is devoted to the analysis and interpretation of these features.

\cite{Muccino2015} establish theoretical predictions concerning short GRBs that originate in a binary NS merger, with and without BH formation. We find that these predictions are fulfilled and that all features are consistent with GRB 090510 resulting from a NS merger leading to the formation of a Kerr-Newman BH.

\subsection{P-GRB}

\begin{figure}
\centering
\includegraphics[angle=180,width=\hsize]{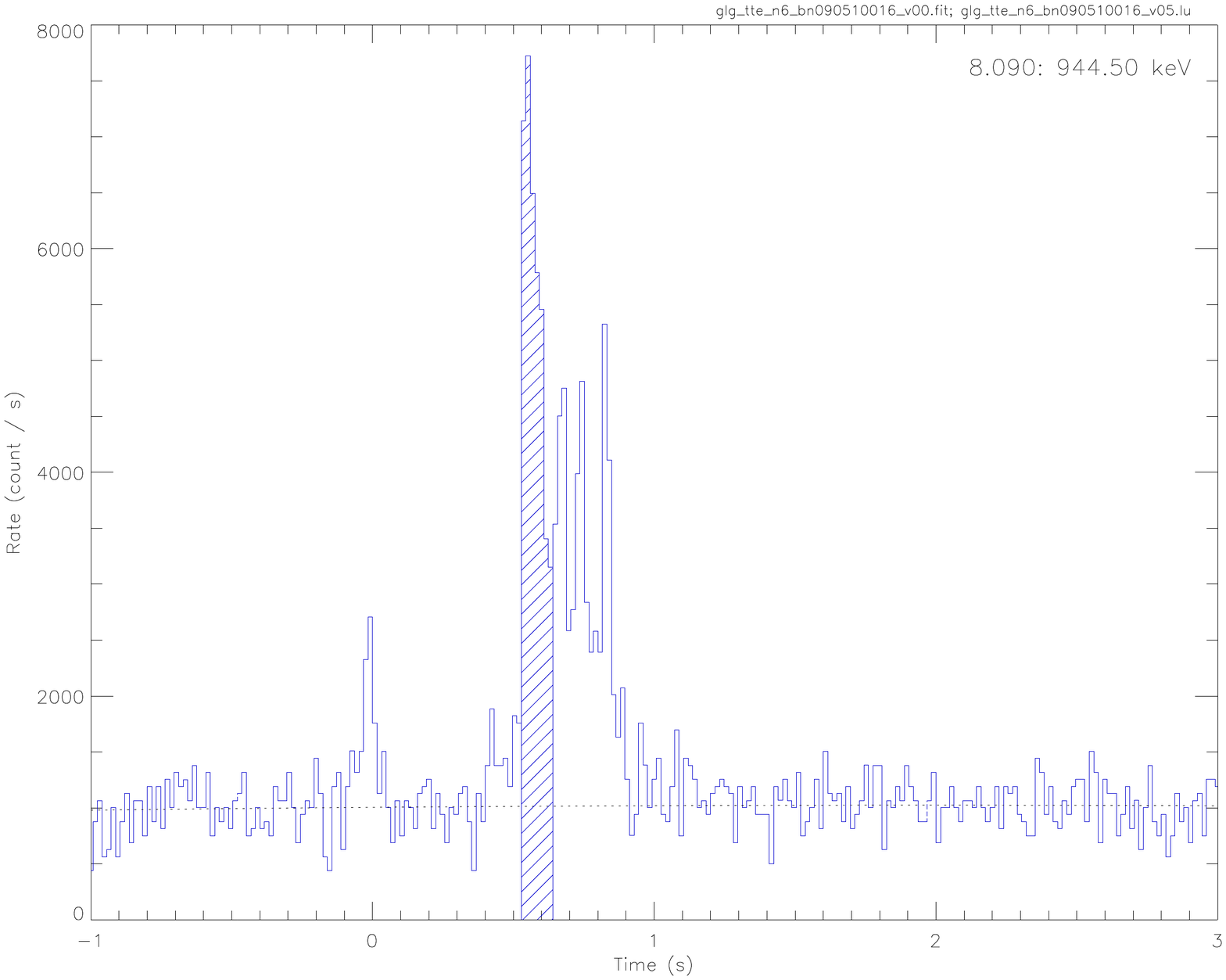}
\caption{Counts light curve of GRB 090510 as seen by the NaI-n6 detector of \emph{Fermi}/GBM with a 16-ms binning. The dashed area represent the interval in which the P-GRB is identified.}
\label{PGRB}
\end{figure}

The identification of the P-GRB is especially relevant to the fireshell analysis, since it marks the reaching of the transparency of the fireshell. The P-GRB is followed by the prompt emission \citep{Ruffini2001c}.
It is suggested in \cite{Muccino2015} that the GeV emission is produced by the newborn BH and starts only after the P-GRB is emitted, at the beginning of the prompt emission.
Here, the bulk of the GeV emission is detected after the first main spike is over. Therefore, we identify the first main spike (from $T_0 + 0.528$ to $T_0 + 0.644$ s, see Figure \ref{PGRB}) with the P-GRB.
The results of the analysis within the fireshell theory (presented hereafter) also offer an \emph{a posteriori} confirmation of this identification of the P-GRB.

\begin{figure}
\centering
\includegraphics[width=0.7\hsize]{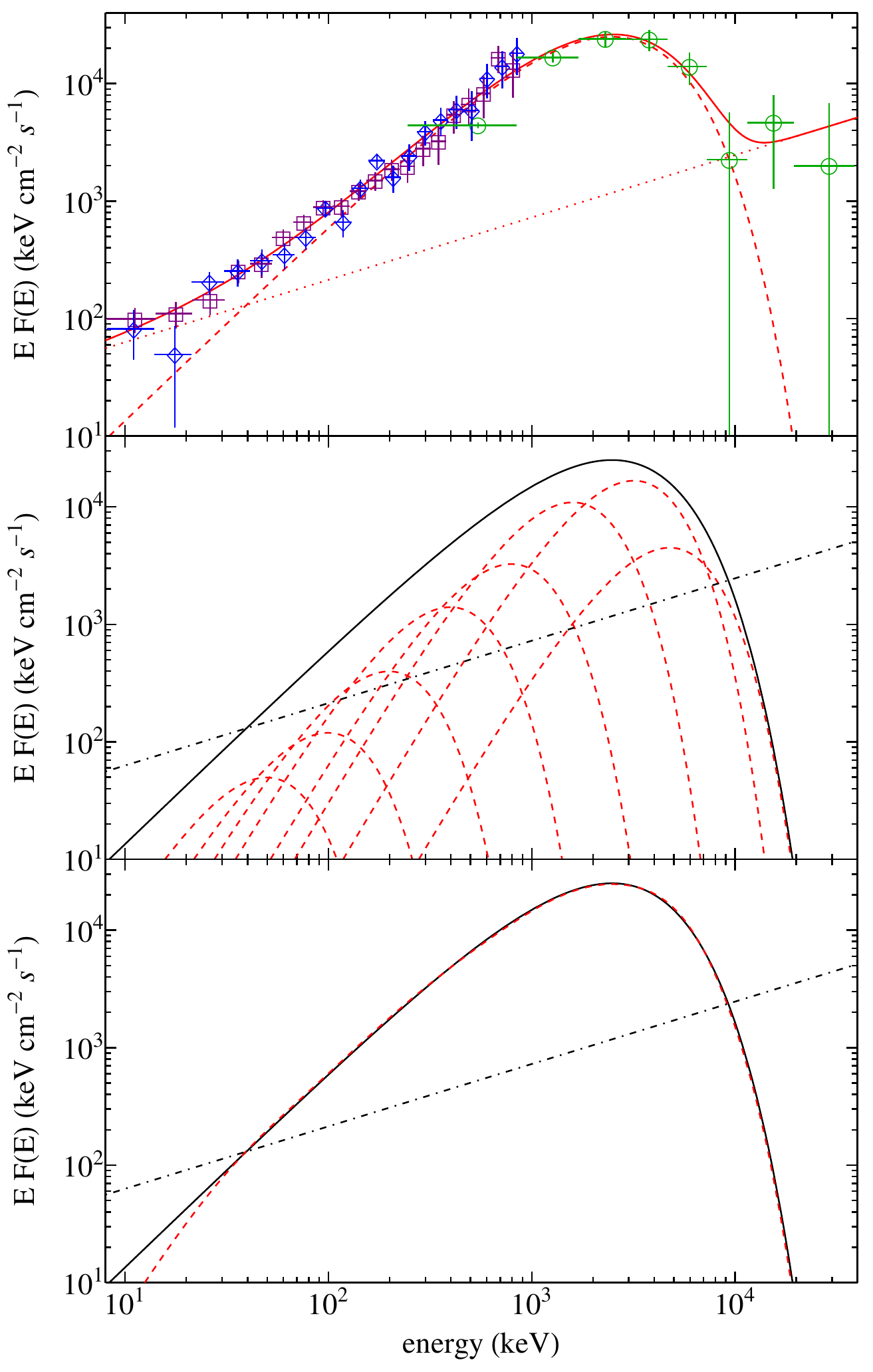}
\caption{Upper panel: the P-GRB spectrum of GRB 090510 from the \textit{Fermi}-GBM NaI-n6 (purple squares) and n7 (blue diamonds), and the BGO-b1 (green circles) detectors, in the time interval from $T_0+0.528$ to $T_0+0.644$ s. The best fit (solid red line) is composed of a power law (dotted red line) and a Comptonized (dashed red curve) models. Middle panel: the above Comptonized model (here the solid black line), viewed as a convolution of thermal components (dashed red curves). The convolution of blackbodies produces the result plotted in the lower panel, namely, a dashed red curve reproducing the Comptonized. The power-law component (dot-dashed black line in the middle and lower panels) is very likely related with a mildly jetted component necessary to fulfill the conservation of the energy and angular momentum of the system.}
\label{comp}
\end{figure}

The best-fit model of the P-GRB spectrum consists of a Comptonized + power law model. We note that a Comptonized component may be viewed as a convolution of blackbodies (see Figure \ref{comp} and Table \ref{tab:compo}, for details).

\begin{table}
\centering
\begin{tabular}{lcc}
\hline\hline
BB & $kT$ (keV)	&  $E_{\rm BB}/E_{\rm P-GRB}$	(\%)\\
 \hline
1   &  $1216$       & $8.8$\\
2   &  $811$   	& $43.6$	\\
3   &  $405$   	& $31.8$\\
4   &  $203$		& $9.6$\\
5   &  $101$ 		& $4.4$\\
6   &  $51$          & $1.2$\\
7   &  $25$          & $0.4$\\
8   &  $13$          & $0.2$\\
 \hline
 \end{tabular}
\caption{The parameters of the blackbody (BB) spectra used in the convolution shown in Figure \ref{comp}. The columns list the number of BB, their temperatures and their energy content with respect to the P-GRB energy computed from the Comptonized model.}
\label{tab:compo}
\end{table}

The geometry of the fireshell is dictated by the geometry of the pair-creation region. It is in general assumed to be a spherically symmetric dyadosphere, which leads to a P-GRB spectrum generally described by a single thermal component in good agreement with the spectral data. \cite{Cherubini2009} found that the region of pair-creation in a Kerr-Newman geometry becomes axially symmetric, thus effectively becoming a dyadotorus. Qualitatively, one expects a pure thermal spectrum resulting from the dyadosphere while a convolution of thermal spectra of different temperatures is expected for a dyadotorus (see Figure \ref{dyadotorus}).

In the present case of GRB 090510, also in view of the good quality of the $\gamma$-ray data, the P-GRB is best fitted by a convolution of thermal spectra. 
The theoretically expected temperatures of the thermal components in the dyadotorus are a function of the polar angle. 
Knowing that the final spectrum at the transparency condition is a convolution of such thermal spectra at different angles, we adopted for simplicity a discrete number of thermal components (see Table \ref{tab:compo}). The number of such thermal components, leading in principle to a continuum, is a function of the quality of the data.
This provides the first indication that indeed the angular momentum plays a role in the merging of the two NSs and that the dyadotorus is formed as theoretically predicted in a series of papers \citep{Cherubini2009,Ruffini2009}. This opens a new area of research which is not going to be addressed in the present article.  
Previous identifications of pure thermal components in the P-GRB of other GRBs (e.g.~\citealt{Izzo2012,Muccino2015}) nevertheless evidence that the angular momentum of the BH formed by GRB 090510 must be substantially large in order to affect the P-GRB spectrum.

Finally, the extra power-law component observed in the P-GRB spectrum is very likely related with a mildly jetted component necessary to fulfill the conservation of the energy and angular momentum of the system.

\begin{figure}
\centering
 \vspace{-0.6cm}
\includegraphics[angle=270,width=0.55\hsize]{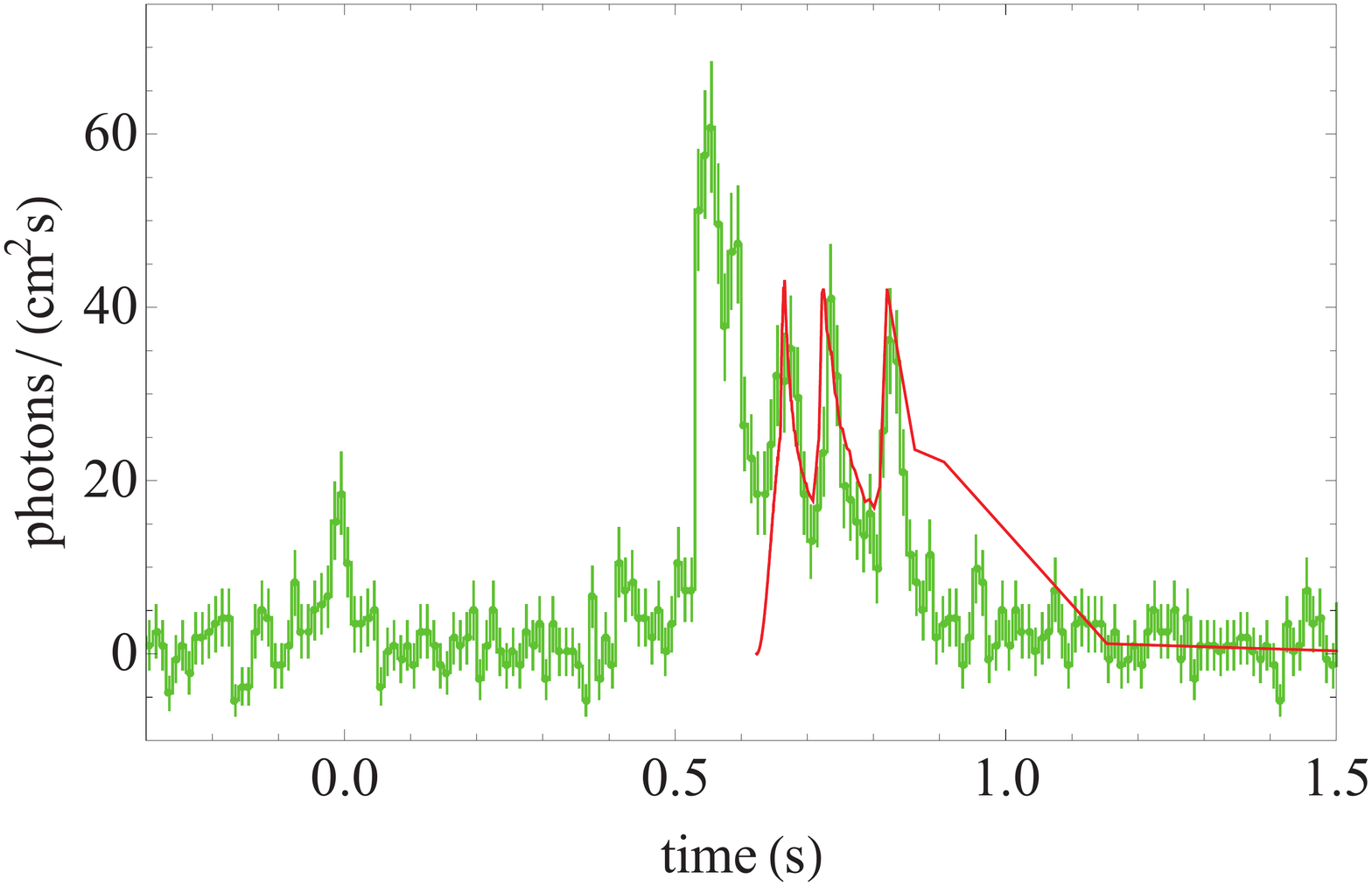}
\\ \vspace{-0.6cm}
\includegraphics[width=0.55\hsize]{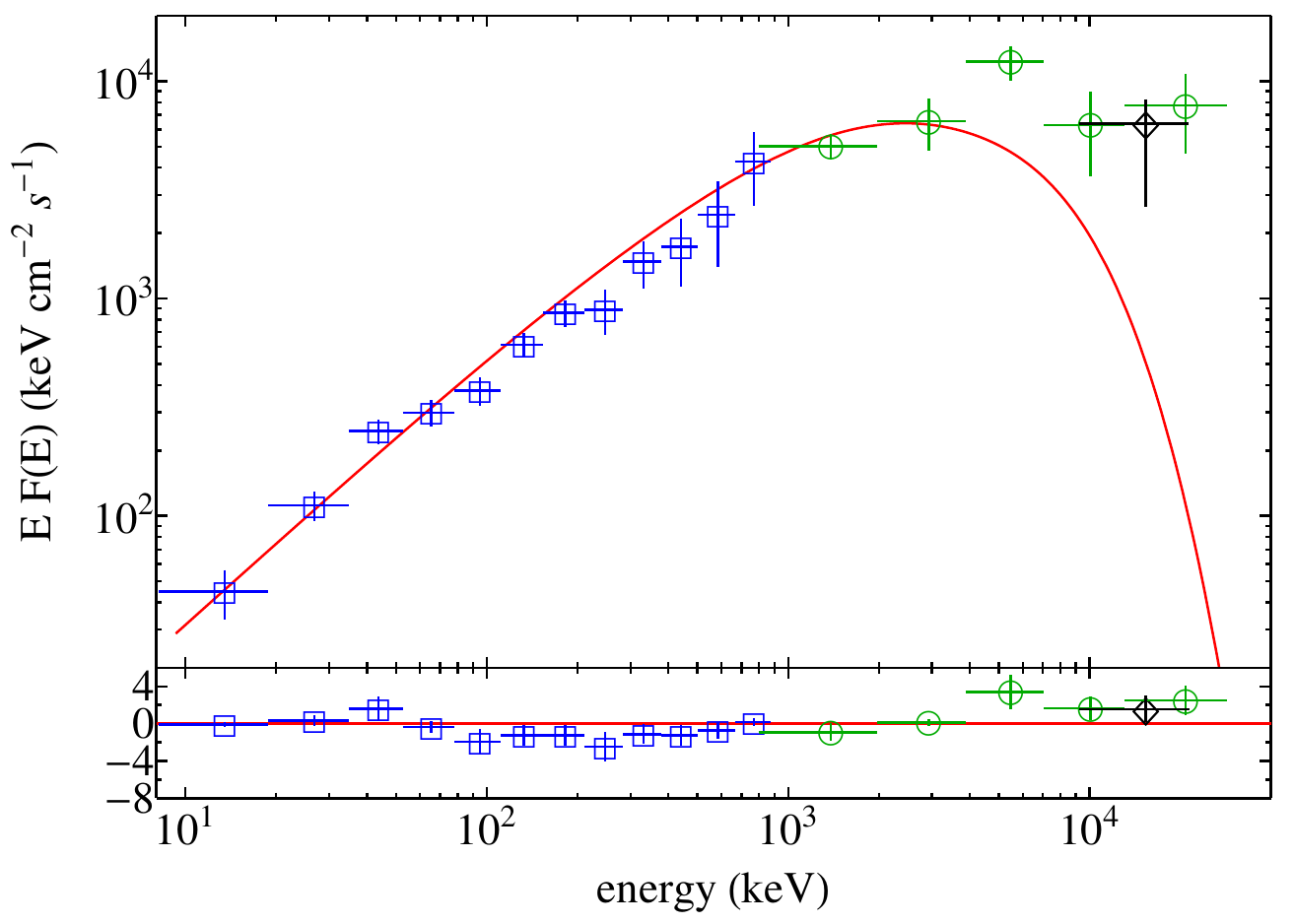}
\\ \vspace{-0.8cm}
\includegraphics[angle=270,width=0.55\hsize]{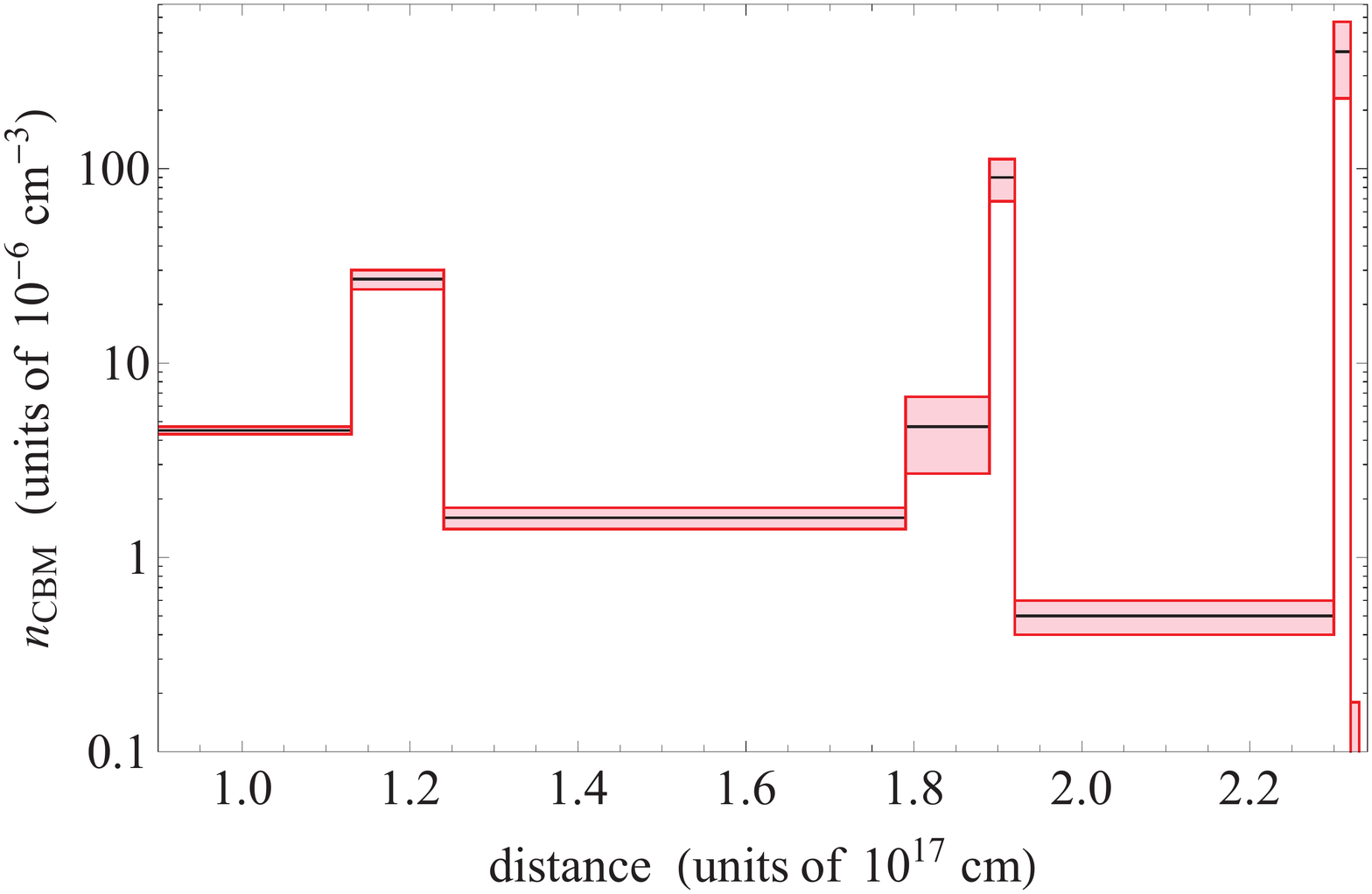}
\\  \vspace{-0.2cm}
\caption{Results of the fireshell simulation of GRB 090510. Upper panel: fit of the prompt emission \textit{Fermi}-GBM NaI-n6 light curve. Middle panel: fit of the corresponding spectrum including the \textit{Fermi}-NaI-6 (blue squares) and BGO-b1 (green circles) data in the time interval from $T_0+0.644$ to $T_0+0.864$ s. A single data point obtained from the Mini-Calorimeter on board AGILE, in the range $10$--$20$ MeV and in the first $0.2$ s of the AGILE light curve (from $T_0+0.5$ to $T_0+0.7$ s in the \textit{Fermi} light curve), is shown for comparison (reproduced from Figure 4 in \citealt{Giuliani2010}). Lower panel: density profile of the CBM inferred from the simulation of CBM clouds of $\sim10^{22}$~g.}
\label{090510simlc}
\end{figure}

\subsection{Prompt emission}

In order to simulate the light curve and spectrum of the prompt emission of GRB 090510, we assume that the initial fireshell energy $E_{e^+e^-}^{\mathrm{tot}}$ is equal to $E_{\mathrm{iso}}$. Since the P-GRB spectrum is not purely thermal, we derive an effective blackbody temperature from the peak energy of the Comptonized component. We obtain a temperature $kT_{\mathrm{obs}} = (633 \pm 62)$ keV.

The fireshell theory allows the determination of all essential quantities of the model from the total pair plasma energy $E_{e^+e^-}^{\mathrm{tot}}$ and from the ratio of the energy contained in the P-GRB to $E_{e^+e^-}^{\mathrm{tot}}$. This ratio directly leads to the baryon load $B$, which in conjunction with $E_{e^+e^-}^{\mathrm{tot}}$ and the relation between the predicted and observed temperatures gives the Lorentz factor at transparency, the temperature of the fireshell at transparency, and the radius at transparency.

Given $E_{\mathrm{iso}} = 3.95 \times 10^{52}$ erg and $E_{\mathrm{P-GRB}} = (42.1 \pm 3.8) \% \  E_{e^+e^-}^{\mathrm{tot}}$, we deduce a baryon load $B = 5.54 \times 10^{-5}$, a Lorentz factor $\gamma = 1.04 \times 10^4$, a temperature at transparency $kT = 1.2$ MeV, and a radius at transparency $r_{\mathrm{tr}} = 7.60 \times 10^{12}$ cm (cf.~Table \ref{fireshellparameters}).

\begin{table}
\centering
\begin{tabular}{lcc}
\hline\hline
  & Parameter & Value \\
 \hline
  & $B$			& $(5.54 \pm 0.70) \times 10^{-5}$ \\
  & $\gamma_{\mathrm{tr}}$ 	& $(1.04 \pm 0.07) \times 10^4$ \\
  &  $r_{\mathrm{tr}}$ 	& $(7.60 \pm 0.50) \times 10^{12}$ cm\\
  &  $E_{e^+e^-}^{\mathrm{tot}}$ & $(3.95 \pm 0.21) \times 10^{52}$ erg \\
  &  $kT_{\mathrm{blue}}$ 	& $(1.20 \pm 0.11) \times 10^3 $ keV\\
  &  $\langle n \rangle$  	& $(8.7 \pm 2.1) \times 10^{-6}$ cm$^{-3}$\\
 \hline
 \end{tabular}
\caption{Parameters derived from the fireshell analysis of GRB 090510: the baryon load $B$, the Lorentz factor at transparency $\gamma_{\mathrm{tr}}$, the fireshell radius at transparency $r_{\mathrm{tr}}$, the total energy of the electron-pair plasma $E_{e^+e^-}^{\mathrm{tot}}$, the blue-shifted temperature of the fireshell at transparency $kT_{\mathrm{blue}}$, and the CBM average density $\langle n \rangle$.}
\label{fireshellparameters}
\end{table}

In order to determine the profile of the CBM, a simulation of the prompt emission following the P-GRB has been performed. The simulation starts at the transparency of the fireshell with the parameters that we determined above. A trial-and-error procedure is undertaken, guided by the necessity to fit the light curve of GRB 090510. The results of this simulation (reproduction of the light curve and spectrum, in the time interval from $T_0+0.644$ to $T_0+0.864$ s, and CBM profile) are shown in Figure \ref{090510simlc}. The average CBM density is found to be $\langle n_{\mathrm{CBM}}\rangle = 8.7 \times 10^{-6}\ \mathrm{cm}^{-3}$. This low value, typical of galactic halo environments, is consistent with the large offset from the center of the host and further justifies the interpretation of GRB 090510 as a short GRB originating in a binary NS merger.

Our theoretical fit of the prompt emission (see red line in the middle plot of Figure \ref{090510simlc}) predicts a cut-off at $\sim10$ MeV. The spectrum at energy $\gtrsim10$ MeV could be affected by the onset of the high energy power-law component manifested both in the data of the Mini-Calorimeter on board AGILE (see top panel of Figure 4 in \citealt{Giuliani2010}) and in the data points from the \textit{Fermi}-GBM BGO-b1 detector.

\subsection{Precursor emission}

There is a weak precursor emission about 0.4 s before the P-GRB (or $\sim 0.21$ s in the cosmological rest frame). 
Two GeV photons have been detected during the precursor emission.
Precursors are commonly seen in long bursts: \cite{Lazzati2005} found that $\sim$ 20\% of them show evidence of an emission preceding the main emission by tens of seconds. Short bursts are less frequently associated with precursors.

No significant emission from the GRB itself is expected prior to the P-GRB -- since it marks the transparency of the fireshell -- but the precursor may be explainable in the context of a binary NS merger by invoking the effects of the interaction between the two NSs just prior to merger. Indeed, it has been suggested that precursor emission in short bursts may be caused by resonant fragmentation of the crusts \citep{Tsang2012} or by the interaction of the NS magnetospheres \citep{HansenLyutikov2001}.

The timescale ($\sim$ 0.21 s between the precursor and the P-GRB)  is consistent with a pre-merger origin of the precursor emission. From its formation to its transparency, the fireshell undergoes a swift evolution. The thermalization of the pair plasma is achieved almost instantaneously ($\sim 10^{-13}$ s, \citealt{Aksenov2007}); and the $e^+ e^-$ plasma of GRB 090510 reaches the ultra-relativistic regime (i.e.~a Lorentz factor $\gamma > 10$) in a matter of $ 4.2 \times 10^{-2}$ s, according to the numerical simulation. The radius of the fireshell at transparency, $r_{\mathrm{tr}} = 7.60 \times 10^{12}$ cm, corresponds to more than a hundred light-seconds; however relativistic motion in the direction of the observer squeezes the light curve by a factor $\sim 2 \gamma^2$, which makes the  fireshell capable of traveling that distance under the observed timescale.

The spectral analysis of this precursor is limited by the low number of counts. \cite{Muccino2013090510} interpreted the spectrum with a blackbody plus power law model. This leads to a blackbody temperature of $34.2 \pm 7.5$ keV. The isotropic energy contained in the precursor amounts to $(2.28 \pm 0.39) \times 10^{51}$ erg.

\subsection{Redshift estimate}

An interesting feature of the fireshell model is the possibility to infer a theoretical redshift from the observations of the P-GRB and the prompt emission. In the case of GRB 090510, a comparison is therefore possible between the measured redshift $z = 0.903 \pm 0.003$ and its theoretical derivation. An agreement between the two values would in particular strengthen the validity of our P-GRB choice, which would in turn strengthen our results obtained with this P-GRB.

The feature of redshift estimate stems from the relations, engraved in the fireshell theory, between different quantities computed at the transparency point: the radius in the laboratory frame, the co-moving frame and blue-shifted temperatures of the plasma, the Lorentz factor, and the fraction of energy radiated in the P-GRB and in the prompt emission as functions of $B$ (see Figure 4 in \citealt{Muccino2015}). Thus, the ratio $E_{\mathrm{P-GRB}} / E_{e^+e^-}^{\mathrm{tot}}$ implies a finite range for the coupled parameters $E_{e^+e^-}^{\mathrm{tot}}$ and $B$ (last panel of Figure 4 in \citealt{Muccino2015}). Assuming $E_{e^+e^-}^{\mathrm{tot}} = E_{\mathrm{iso}}$, this ratio is known since it is equal to the ratio between the observed fluences of the respective quantities:
\begin{equation}
\frac{E_{\mathrm{P-GRB}} }{ E_{e^+e^-}^{\mathrm{tot}}} \approx \frac{4\pi S_{\mathrm{P-GRB}} d_l^2(z) / (1+z)}{4\pi S_{e^+e^-}^{\mathrm{tot}} d_l^2(z) / (1+z)} = \frac{S_{\mathrm{P-GRB}} }{S_{e^+e^-}^{\mathrm{tot}}}
\end{equation}

With the measured values $S_{\mathrm{P-GRB}} = (9.31 \pm 0.76) \times 10^{-6}$ erg cm$^{-2}$ and $S_{e^+e^-}^{\mathrm{tot}} = (2.19 \pm 0.18) \times 10^{-5}$ erg cm$^{-2}$, we find $E_{\mathrm{P-GRB}} / E_{e^+e^-}^{\mathrm{tot}} = (42.1 \pm 3.8)\%$.

In addition, knowing the couple [$E_{e^+e^-}^{tot}$, $B$] gives the (blue-shifted towards the observer) temperature of the fireshell at transparency $kT_{\mathrm{blue}}$ (Figure 4 in \citealt{Muccino2015}, second panel). But we also have the following relation between $kT_{\mathrm{blue}}$ and the observed temperature at transparency $kT_{\mathrm{obs}}$, linking their ratio to the redshift:
\begin{equation}
\frac{kT_{\mathrm{blue}}}{kT_{\mathrm{obs}}} = 1 + z\ .
\end{equation}

Finally, since we assume that $E_{e^+e^-}^{\mathrm{tot}} = E_{\mathrm{iso}}$, we also have an expression of $E_{e^+e^-}^{\mathrm{tot}}$ as a function of $z$ using the formula of the K-corrected isotropic energy:
\begin{equation}
E_{\mathrm{iso}} = 4\pi d_l^2(z) \frac{S_{\mathrm{tot}}}{1+z} \frac{\int_{1/(1+z)\ \mathrm{keV}}^{10000/(1+z)\ \mathrm{keV}} EN(E) dE}{\int_{8\ \mathrm{keV}}^{40000\ \mathrm{keV}} EN(E) dE}
\end{equation}

\noindent where $N(E)$ is the photon spectrum of the GRB and the fluence $S_{\mathrm{tot}}$ is obtained in the full GBM energy range 8 -- 40000 keV.

The use of all these relations allows a redshift to be determined by an iterative procedure, testing at every step the value of the parameters $E_{e^+e^-}^{\mathrm{tot}} (z)$ and $kT_{\mathrm{blue}}$. The procedure successfully ends when both values are consistent according to the relations described above. In the case of GRB 090510, we find $z = 0.75 \pm 0.17$, which provides a satisfactory agreement with the measured value $z = 0.903 \pm 0.003$.

\subsection{GeV emission}\label{GeVemission}

GRB 090510 is associated with a high-energy emission, consistently with all other observed S-GRBs, i.e. energetic events with $E_{\mathrm{iso}} \gtrsim 10^{52}$ erg. The only case of a S-GRB without GeV emission, namely GRB 090227B, has been explained by the absence of alignment between the LAT and the source at the time of the GRB emission. Nevertheless evidence of some GeV emission in this source has been recently obtained (Ruffini et al., in preparation).

\begin{figure*}
\centering
\includegraphics[width=0.8\hsize]{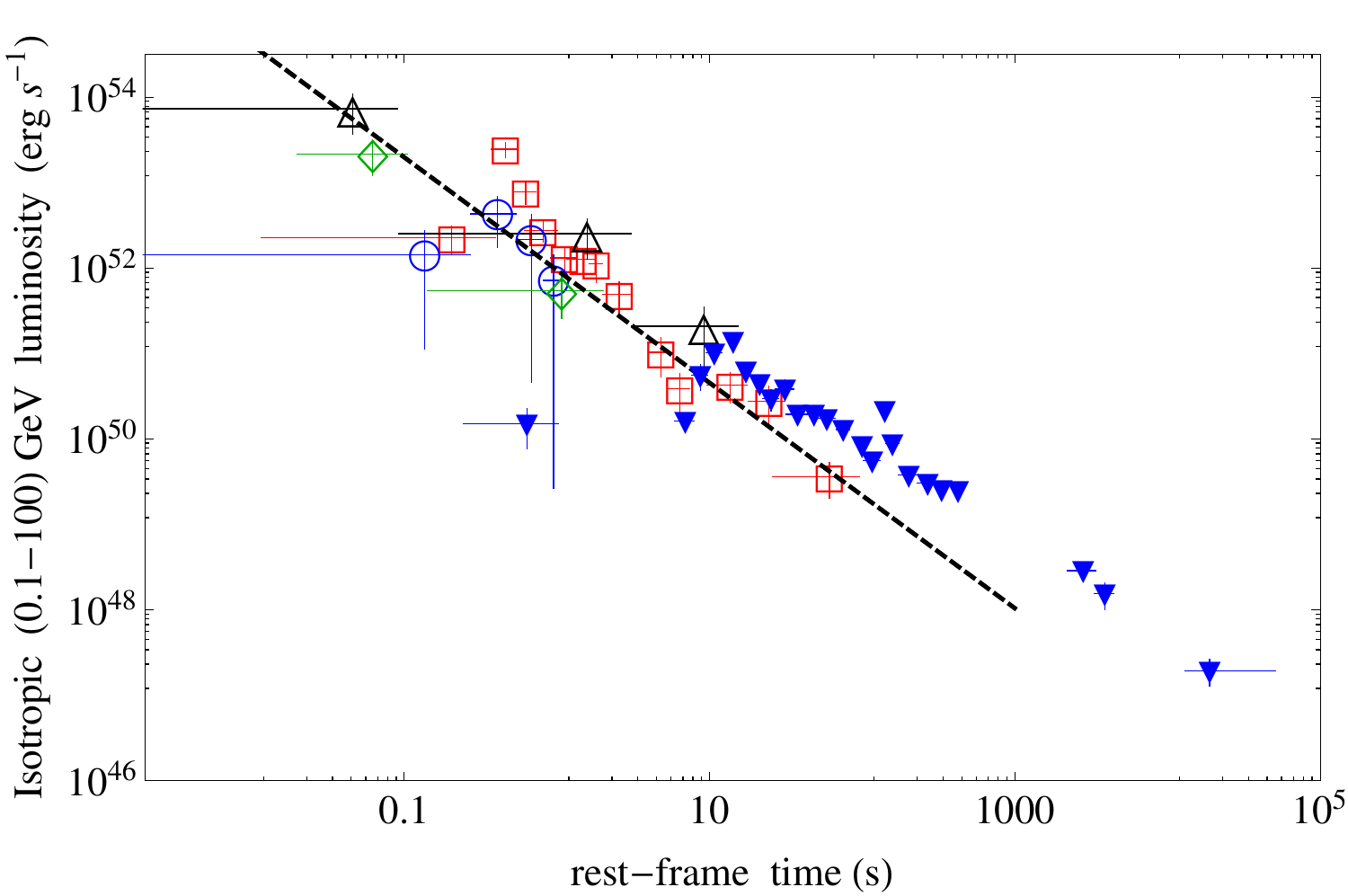}
\caption{Isotropic rest-frame $0.1$--$100$ GeV luminosity light curves of the S-GRBs 090510 (red squares), 081024B (green diamonds), 140402A (black triangles), and 140619B (blue circles) compared to that of the BdHNe 130427A (blue reverse triangles). The dashed black line marks the common behavior of all the S-GRB light curves which goes as $t^{-1.32}$. In our approach this communality follows straightforwardly from the equality of the masses of the emerging extreme BH.}
\label{090510GeV}
\end{figure*}

The GeV light curve of GRB 090510 is plotted in Figure \ref{090510GeV} together with other S-GRB light curves and showing a common power-law behavior, which goes as $t^{-1.32}$, similar to the clustering of the GeV light curves found by \citet{2014MNRAS.443.3578N}. These S-GRBs are compared with that of the BdHNe 130427A which shares a similar behavior. 
\cite{Muccino2015} suggest and argue that the GeV emission is related to the presence of a BH and its activity. This view is supported by the fact that the GeV emission is delayed with respect to the $\gamma$-ray emission: it starts only after the P-GRB is over.

The GeV emission of GRB 090510 is particularly intense, reaching $E_{\mathrm{LAT}}=(5.78\pm0.60)\times10^{52}$ erg. Such a large value, one of the largest observed among S-GRBs, is consistent with the large angular momentum of the newborn BH.
This energetic cannot be explained in terms of NSs in view of the lower value of the gravitational binding energy.

\begin{figure}
\centering
\includegraphics[width=\hsize]{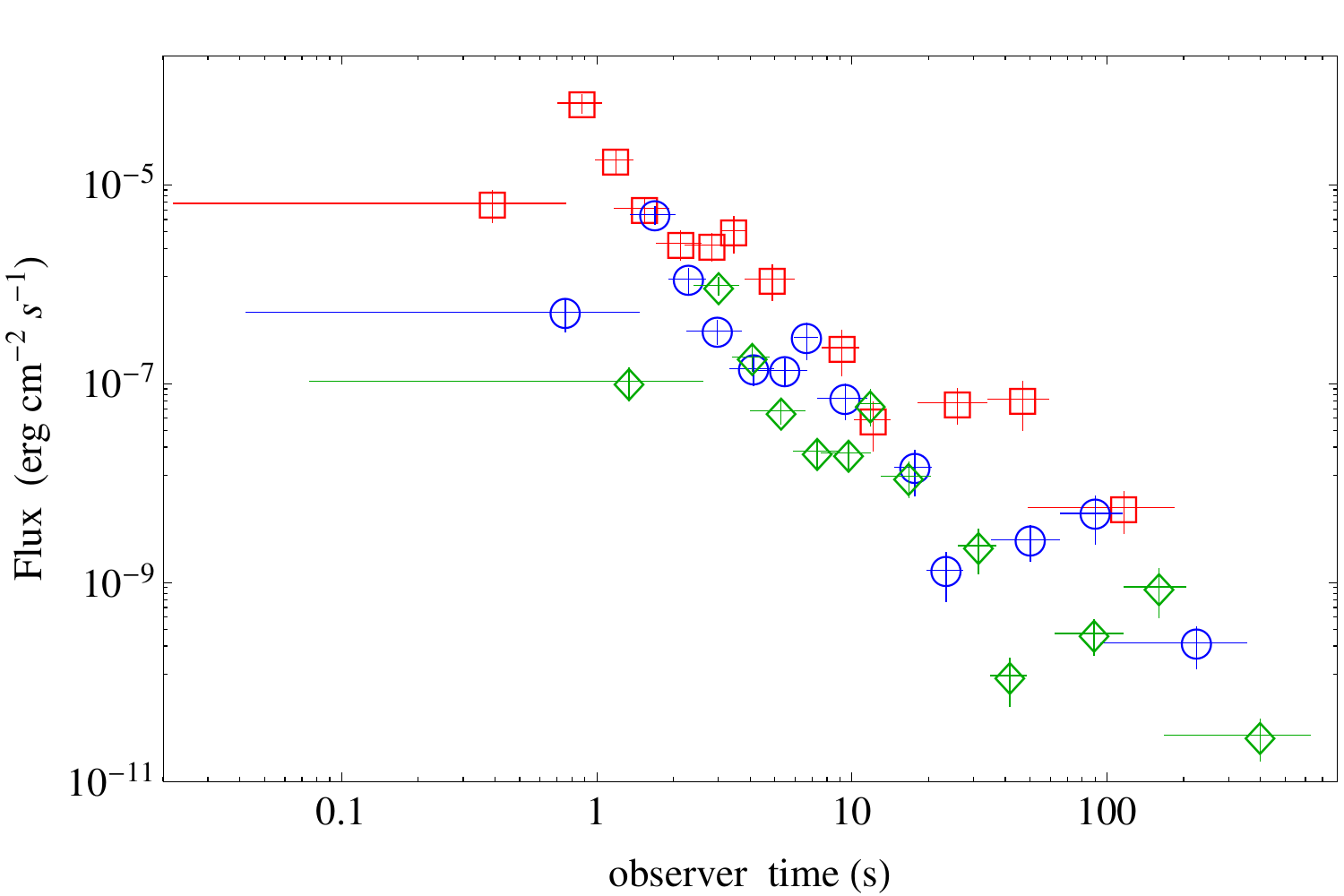}
\caption{The observed $0.1$--$100$ GeV flux light curve of the S-GRBs 090510 (red squares), and the corresponding ones obtained by translating this S-GRB at $z=2.67$ (blue circles) and at $z=5.52$ (green diamonds).}
\label{GeVz}
\end{figure}

The absence of GeV emission in S-GRFs is also confirmed from the strong upper limit to the GeV emission for S-GRBs imposed by the Fermi-LAT sensitivity. 
We assume for a moment that the GeV emission of a S-GRF is similar to that of S-GRB. We then compute the observed GeV flux light curve of S-GRB 090510 at different redshifts, e.g., $z=2.67$ and $5.52$, which correspond to the redshifts of the S-GRB 081024B and of the S-GRB 140402A, respectively (Aimuratov et al., in preparation). 
The result is that if we compare these computed flux light curves with the \textit{Fermi}-LAT sensitivity of the Pass 8 Release 2 Version 6 Instrument Response Functions, 
\footnote{\url{http://www.slac.stanford.edu/exp/glast/groups/canda/lat_Performance_files/broadband_flux_sensitivity_p8r2_source_v6_all_10yr_zmax100_n03.0_e1.50_ts25.png}} which is approximately $10^{-11}$ erg cm$^{-2}$s$^{-1}$, all of them are always well above the LAT broadband sensitivity by a factor $\sim10^5$ (see Figure \ref{GeVz}). 
This result does not depend on the choice of the source. 
In their rest-frame all the S-GRB GeV light curves follow a similar behavior. 
Therefore, the GeV emission of S-GRB 090510 is always above $\sim10^5$ times to the LAT sensitivity, even at higher redshifts. 
If we now assume that S-GRFs do conform to the same behavior of S-GRBs, the absence of detection of GeV emission implies that the S-GRFs have necessarily fluxes at least $10^5$--$10^6$ times smaller than those of S-GRBs.

\subsection{On the energy requirement of the GeV emission}\label{reqGeVemission}

In order to estimate the energy requirement of the $0.1$--$100$ GeV emission of Figure \ref{090510GeV} we consider the accretion of mass $M_{\rm acc}$ onto a Kerr-Newman BH, dominated by its angular momentum and endowed with electromagnetic fields not influencing the geometry, which remains approximately that of a Kerr BH. We recall that if the infalling accreted material is in an orbit co-rotating with the BH spin, up to $\eta_+=42.3\%$ of the initial mass is converted into radiation, for a maximally rotating Kerr BH, while this efficiency drops to $\eta_-=3.8\%$, when the infalling material is on a counter-rotating orbit (see Ruffini \& Wheeler 1969, in problem 2 of $\S$ 104 in \citealt{LL2003}).
Therefore, the GeV emission can be expressed as 
\begin{equation}
\label{accretion}
E_{\rm LAT}= f_b^{-1}\eta_\pm M_{\rm acc} c^2\ ,
\end{equation}
and depends not only on the efficiency $\eta_\pm$ in the accretion process of matter $M_{\rm acc}$, but also on the geometry of the emission described by the beaming factor $f_b\equiv1-\cos\theta$ (here $\theta$ is the half opening angle of jet-like emission).

Depending on the assumptions we introduce in Equation \ref{accretion}, we can give constraints on the amount of accreted matter or on the geometry of the system.
 
For an isotropic emission, $f_b\equiv1$, the accretion of $M_{\rm acc}\gtrsim0.08$~M$_\odot$, for the co-rotating case, and of $M_{\rm acc}\gtrsim0.86$~M$_\odot$, for the counter-rotating case, is required.

Alternatively, we can assume that the accreted matter comes from the crustal material from an $1.6+1.6$~M$_\odot$ NS--NS binary progenitor. The crustal mass from the NL3 nuclear model for each of these NSs is $M_c=4.30\times10^{-5}$~M$_\odot$ \citep[see, e.g.,][and figure \ref{Nanda}]{Belvedere2012}. Assuming that crustal material accounts also for the baryon load mass, e.g., $M_B\equiv E^{tot}_{e^+e^-}B/c^2=1.22\times10^{-6}$ M$_\odot$, the total available mass for accretion is $M_{\rm acc}\equiv 2M_c-M_B=8.48\times10^{-5}$ M$_\odot$. Then, the presence of a beaming is necessary: from Equation \ref{accretion}, a half opening beaming angle $\theta\gtrsim2^{\rm o}.70$, for co-rotating case, and $\theta\gtrsim0^{\rm o}.81$, for the counter-rotating case, would be required.

The above considerations are clearly independent from the relativistic beaming angle $\theta_r=\gamma_{\rm LAT}^{-1}\approx0^{\rm o}.1$, where the lower limit on the Lorentz factor $\gamma_{\rm LAT}\approx550$ has been derived, in a different context, by \citet{Lithwick} to the GeV luminosity light curve (see Figure \ref{090510GeV}).

Further consequences on these results for the estimate of the rate of these S-GRBs will be presented elsewhere (Ruffini et al. in preparation).

\section{Conclusions}

It is interesting to recall some of the main novelties introduced in this paper with respect to previous works on GRB 090510. 
Particularly noteworthy are the differences from the previous review of short bursts by \citet{2007PhR...442..166N}, made possible by the discovery of the high energy emission by the Fermi team in this specific source \citep{Ackermann2010}.
A new family of short bursts characterized by the presence of a BH and associated high energy emission when LAT data are now available, comprehends GRBs 081024B, 090227B, 090510, 140402A, and 140619B (see, e.g., Figure \ref{090510GeV}). 
The excellent data obtained by the Fermi team and interpreted within the fireshell model has allowed to relate in this paper the starting point of the high energy emission with the birth of a BH.

Our fireshell analysis assumes that the $\gamma$-ray and the GeV components originate from different physical processes. First, the interpretation of the prompt emission differs from the standard synchrotron model: we model the collisions of the baryon accelerated by the GRB outflow with the ambient medium following a fully relativistic approach (see Section 2). Second, we assume that the GeV emission originates from the matter accretion onto the newly-born BH and we show that indeed the energy requirement is fulfilled. This approach explains also the delayed onset of the GeV emission, i.e., it is observable only after the transparency condition, namely after the P-GRB emission.

The joint utilization of the excellent data from the \textit{Fermi}-GBM NaI-n6 and n7 and the BGO-b1 detectors and from the Mini-Calorimeter on board AGILE \citep{Giuliani2010} has given strong observational support to our theoretical work.
GRB 090510 has been analyzed in light of the recent progress achieved in the fireshell theory and the resulting new classification of GRBs. 
We show that GRB 090510 is a S-GRB, originating in a binary NS merger (see figure~\ref{rt_gw}). Such systems, by the absence of the associated SN events, are by far the simplest GRBs to be analyzed. 
Our analysis indicates the presence of three distinct episodes in S-GRBs: the P-GRB, the prompt emission, and the GeV emission.
By following the precise identification of successive events predicted by the fireshell theory, we evidence for the first indication of a Kerr BH or, possibly, a Kerr-Newman BH formation:
\begin{itemize}
\item The P-GRB spectrum of GRB 090510, in the time interval from $T_0+0.528$ to $T_0+0.644$ s, is best-fitted by a Comptonized component (see figures~\ref{PGRB} and \ref{comp} and table~\ref{tab:fit}), which is interpreted as a convolution of thermal spectra originating in a dyadotorus (see \citealt{Cherubini2009} and \citealt{Ruffini2009}, figure~\ref{dyadotorus}, and section 2).
\item The prompt emission follows at the end of the P-GRB (see figure~\ref{spectotal}). The analysis of the prompt emission within the fireshell model allows to determine the inhomogeneities in the CBM giving rise to the spiky structure of the prompt emission and to estimate as well an averaged CBM density of $\langle n_{\rm CBM}\rangle=8.7\times10^{-6}$~cm$^{-3}$ obtained from a few CBM clouds of mass $\sim10^{22}$ g and typical dimensions of $\sim10^{16}$ cm (see figure~\ref{090510simlc}). Such a density is typical of galactic halos where binary NS are expected to migrate due to large natal kicks.
\item The late X-ray emission of GRB 090510 does not follow the characteristic patterns expected in BdHN events (see figure~\ref{episode3} and \citealt{Pisani2013}).
\item The GeV emission occurs at the end of the P-GRB emission and is initially concurrent with the prompt emission. This sequence occurs in both S-GRBs \citep{Muccino2015} and BdHNe \citep{Wang2015}. This delayed long lasting ($\approx200$ s) GeV emission in GRB 090510 is one of the most intense ever observed in any GRB \citep[see figure~\ref{090510GeV} and][]{Ackermann2013,Ruffini2016}.
\item We then consider accretion on co-rotating and counter-rotating orbits (see Ruffini \& Wheeler 1969, in problem 2 of $\S$ 104 in \citealt{LL2003}) around an extreme Kerr BH. Assuming the accretion of the crustal mass $2M_c=8.60\times10^{-5}$~M$_\odot$ from a $1.6+1.6$~M$_\odot$ NS--NS binary, fulfilling global charge neutrality (see figure~\ref{Nanda}), geometrical beaming angles of $\theta\gtrsim0^{\rm o}.81$, for co-rotating case, and $\theta\gtrsim2^{\rm o}.70$, for the counter-rotating case, are inferred. In order to fulfill the transparency condition, the initial Lorentz factor of the jetted material has to be $\gamma\gtrsim550$ (see section 6.6).
\item While there is evidence that the GeV emission must be jetted, no beaming appears to be present in the P-GRB and in the prompt emission, with important consequence for the estimate of the rate of such events \citep{Ruffini2016}.
\item The energetic and the possible beaming of the GeV emission requires the presence of a Kerr BH, or a Kerr-Newman BH dominated by its angular momentum and with electromagnetic fields not influencing the geometry (see also section 6.5).
\item The self-consistency of the entire procedure has been verified by estimating, on the ground of the fireshell theory, the cosmological redshift of the source. The theoretical redshift is $z=0.75\pm0.17$ (see section~6.4), close to and consistent with the spectroscopically measured value $z=0.903\pm0.003$ \citep{Rau2009}.
\item The values of $E_{\mathrm{peak}}$ and $E_{\mathrm{iso}}$ of GRB 090510 fulfill with excellent agreement the MuRuWaZha relation \citep[see section~5.2, figure~\ref{calderone} and][]{RuffiniGCN}.
\end{itemize}

The main result of this article is that the dyadotorus manifests itself by the P-GRB emission and clearly preceeds the prompt emission phase, as well as the GeV emission originating from the newly-formed BH. This contrasts with the usual assumption made in almost the totality of works relating BHs and GRBs in which the BH preceeds the GRB emission. 
In conclusion, in this article, we take GRB 090510 as the prototype of S-GRBs and perform a new time-resoved spectral analysis, in excellent agreement with that performed by the AGILE and the \textit{Fermi} teams. Now this analysis, guided by a theoretical approach successfully tested in this new family of S-GRBs, is directed to identify a precise sequence of different events made possible by the exceptional quality of the data of GRB 090510. This include a new structure in the thermal emission of the P-GRB emission, followed by the onset of the GeV emission linked to the BH formation, allowing, as well, to derive the strucutre of the circumburst medium from the spiky structure of the prompt emission. This sequence, for the first time, illustrates the formation process of a BH.

It is expected that this very unique condition of generating a jetted GeV emission in such a well defined scenario of a newly-born BH will possibly lead to a deeper understanding of the equally jetted GeV emission observed, but not yet explained, in a variety of systems harboring a Kerr BH. 
Among these systems we recall binary X-ray sources \citep[see, e.g.,][and references therein]{1978pans.proc.....G}, microquasars \citep[see, e.g.,][and references therein]{2015IAUS..313..370M}, as well as, at larger scale, active galactic nuclei \citep[see e.g.,][and references therein]{2015A&A...579A..34A}.

\acknowledgments

We thank the Editor and the Referee for their comments which helped to improve the presentation and the contextualization of our results.
We are indebted to Marco Tavani for very interesting comments, as well as for giving us observational supporting evidences.
This work made use of data supplied by the UK \emph{Swift} data Center at the university of Leicester. 
M. E., M. K., and Y. A. are supported by the Erasmus Mundus Joint Doctorate Program by Grant Numbers 2012-1710, 2013-1471, and 2014-0707 respectively, from the EACEA of the European Commission. 
C.C. acknowledges INdAM-GNFM for support.
M.M. acknowledges the partial support of the project N 3101/GF4 IPC-11, and the target program F.0679 of the Ministry of Education and Science of the Republic of Kazakhstan.

\end{document}